\documentclass{article}

\usepackage{arxiv}

\usepackage[utf8]{inputenc} 
\usepackage[T1]{fontenc}    
\usepackage{hyperref}       
\usepackage{url}            
\usepackage{booktabs}       
\usepackage{amsfonts}       
\usepackage{nicefrac}       
\usepackage{microtype}      
\usepackage{lipsum}		
\usepackage{graphicx}
\usepackage{natbib}
\usepackage{doi}
\usepackage{multirow}
\usepackage{amsmath}
\usepackage{listings}
\usepackage{float}
\usepackage{xcolor}

\title{All you need for horizontal slicing in 5G network }

\author{ \href{https://orcid.org/0000-0002-9532-2453}{\includegraphics[scale=0.06]{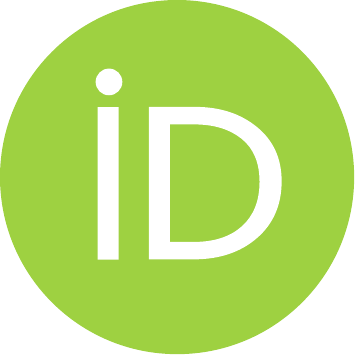}\hspace{1mm}Hamza Kheddar}\thanks{Use footnote for providing further
		information about author (webpage, alternative
		address)---\emph{not} for acknowledging funding agencies.} \\
	Department of Electrical engineering\\
	University of Medea\\
	Medea, 26000, Algeria \\
	\texttt{kheddar.hamza@univ-medea.dz} \\
	\And
	{\hspace{1mm}Soufiane Ouldkhaoua} \\
	Department of Electrical engineering\\
	University of Medea\\
	Medea, 26000, Algeria \\
	\And
	{\hspace{1mm} Riadh Bouguerra} \\
	Department of Electrical engineering\\
	University of Medea\\
	Medea, 26000, Algeria \\
}	

\renewcommand{\shorttitle}{All you need for horizontal slicing in 5G network}

\hypersetup{
pdftitle={A template for the arxiv style},
pdfsubject={q-bio.NC, q-bio.QM},
pdfauthor={David S.~Hippocampus, Elias D.~Striatum},
pdfkeywords={First keyword, Second keyword, More},
}

\begin{document}
\maketitle

\begin{abstract}
The telecommunication field has seen unprecedented growth in the last decade that has led to the release of several generations that have been committed to satisfy users by increasing the data rate and reducing the latency, especially in the 5G network.
With fully commercialized 5G networks that is already launched in many country, Software-defined network (SDN) and network function virtualization (NFV) will facilitate the implementation of NS. SDN and NFV will serve as the basis for NS, allowing efficient use of both physical and virtual resources. This paper makes it possible to analyze, propose an efficient model, and utilize all of the available resources of the 5G network.

\end{abstract}

\keywords{5G network \and Software-defined network  \and Network function virtualization \and Network slicing }

\section{Introduction}
\label{sec:1}
Telecommunications companies and technology businesses all around the world have been collaborating to explore and create innovative technological solutions in order to address rising consumer and industrial demand for mobile data. Fifth-generation (5G) mobile technologies are the next generation of mobile communications technologies aimed at improving current mobile networks. 
Faster speeds, more capacity, and the ability to support new features and services are all expected from 5G networks. The latter technologies were created in response to the growing need for mobile data (i.e., more people using more data on more devices). 5G cellular networks are intended to provide throughput of up to 10 Gbps, ten times energy efficiency, and high reliability, and latency up to 1 ms, one thousand greater than usual number of devices.
5G networks are aimed for providing three types of services which are: massive machine type communication (mMTC), ultra-reliable low latency communication (URLLC), and enhanced mobile broadband (eMBB). Network slicing (NS) is one of the key technologies in the 5G architecture that has the ability to divide the physical network into multiple logical networks, such as slices, with different network characteristics. It's a useful option for enabling 5G networks to provide a wide range of services and applications, including augmented reality, e-health care, mobile gaming, smart banking, smart farming, and smart transportation systems. 
This paper contains an overview on the 5G network with its architecture and different characteristics, the new features and function that differentiate from the previous generations.
A practical part about 5G horizontal slicing (HS) will be simulated, to extract users’ weights for each service, in order to satisfy service layer agreements (SLAs) between the infrastructure providers (telecommunication operator) and tenants. 

\section{Background}
\subsection{5G services and requirements}
\label{sec:2}
The international telecommunication union (ITU) has classified 5G mobile network services into three categories \cite{marsh20185g}: 
\begin{itemize}
    \item \textbf{eMBB: } Refers to a straighter forward progression of today's mobile broadband services, allowing huge data amounts and a better user experience, such as supporting even higher end-user data rates, ultra-high definition (UHD) videos, virtual reality (VR), and augmented reality (AR).
   \item \textbf{mMTC:} This term refers to services with a large number of devices, such as remote sensors, equipment monitoring, and controllers. Low device cost and low device energy consumption are key needs for such services, allowing for at least several years of device battery life. Because each device typically consumes low energy and produces a limited quantity of data, high data rate support is less important. 
   \item \textbf{URLLC:} Aims to meet expectations for the demanding digital industry and focuses on low latency and extremely high reliability services, such as assisted and automated driving, and factory automation. 
\end{itemize}

\begin{figure}
	\centering
	\includegraphics[]{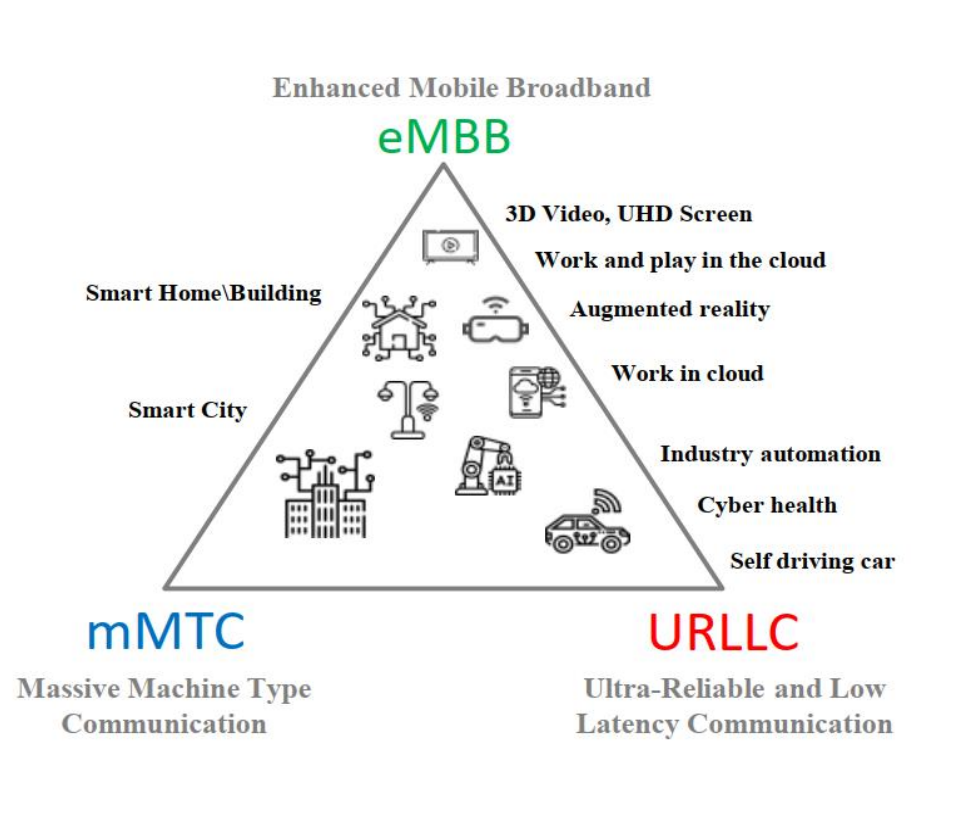}
	\caption{Services categories provided in 5G.}
	\label{fig1}
\end{figure}

Figure \ref{fig1} depicts the different services that the mMTC, URLLC and eMBB offers in the 5G network. 

By accounting for the majority of needs, the following set of 5G requirements is gaining industry acceptance. Table \ref{tab1} summarrizes the quantitative terms of the qualitative requirements of a 5G mobile network.

\begin{table}[]
	\caption{5G mobile network requirements.}
	\label{tab1}
\begin{tabular}{p{3.5cm}p{9cm}p{2.5cm}}
\hline
 Capability& Description  & 5G requirement \\
 \hline
 Downlink peak data rate & \multirow{2}{*}{Minimum maximum data rate technology must support} & 20 Gbps  \\
 \cline{1-1}
 \cline{3-3}
 Uplink peak data rate &                   & 10 Gbps \\
 \hline
Data rate in up-link & \multirow{2}{*}{In a densely populated urban, test setting for data rate is 95\% of } & 50 Mbps  \\
 \cline{1-1}
 \cline{3-3}
Data rate in down-link &    the time               & 100 Mbps \\
 \hline
\multirow{2}{*}{Latency} & \multirow{2}{*}{Contribution of the radio network to packet travel time} & 1 ms (uplink) \\
 \cline{3-3}
 &                   & 4 ms (Downlink) \\
 \hline
 Mobility &Maximum speed for handoff and QoS requirements&500 km/h\\
 \hline
 Density of connections&Total number of devices per unit area& 10$^{6}$/km$^2$\\
 \hline
 Efficiencies in energy &Per unit of energy consumption, data is received or sent (by device or network)& Same as 4G\\
 \hline
 Traffic capacity in area &Traffic totals for the whole service area& 10 Mbps/m$^2$\\
 \hline
Peak downlink spectrum efficiency &Throughput per unit wireless bandwidth and per network cell&30 bps/Hz\\
 \hline
\end{tabular}
\end{table}

\subsection{Software-defined networking and network function virtualization in 5G }
\label{sec:3}

\subsubsection{Software defined networking }
\begin{figure}
	\centering
	\includegraphics[]{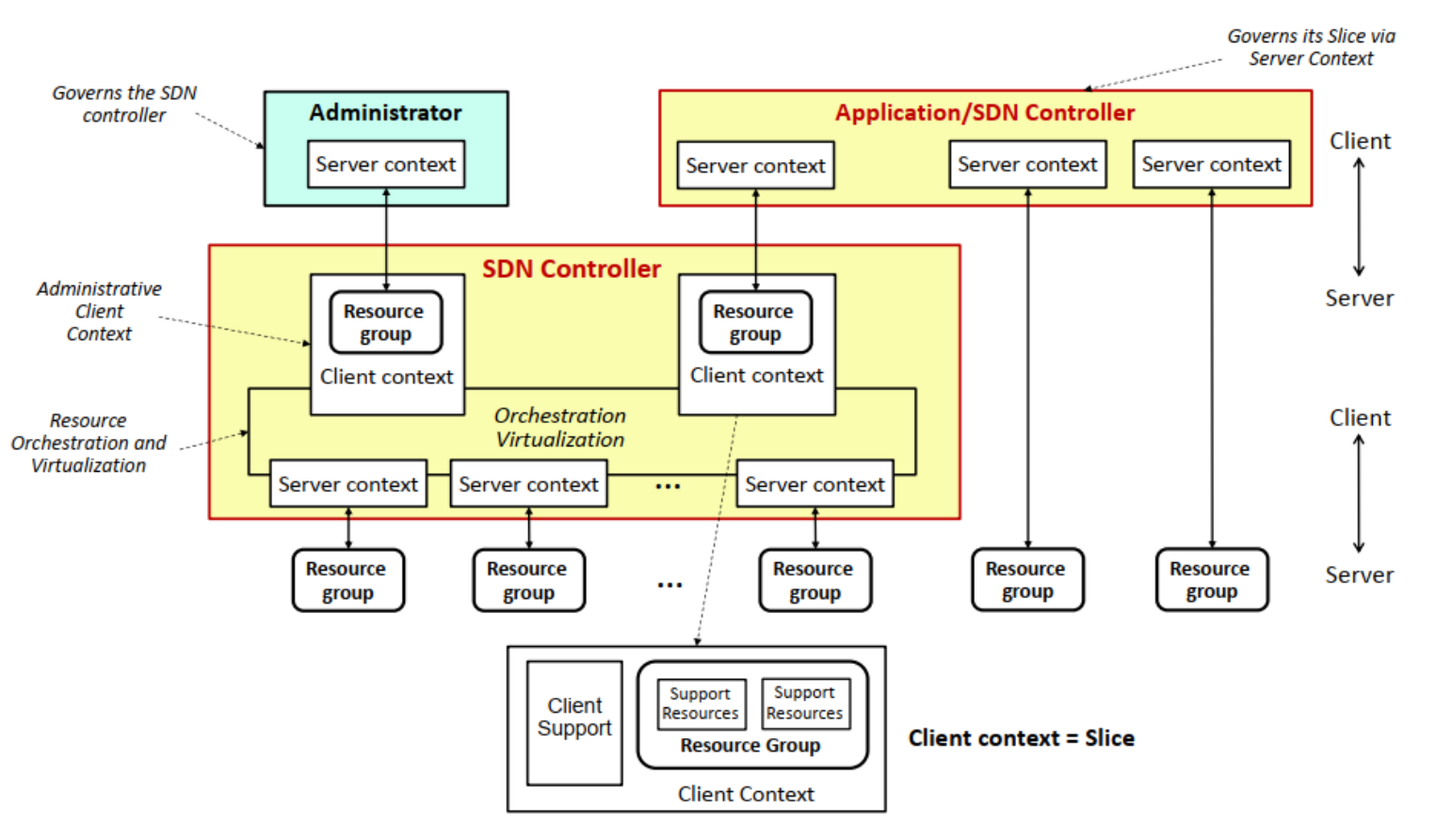}
	\caption{SDN slicing network architecture.}
	\label{fig2}
\end{figure}

The goal of software defined networking (SDN) is to increase networks' flexibility and adaptability. SDN was initially characterized as a method of planning, constructing, and administering networks that divides the network's control and forwarding planes, it allows the network's control plane to be directly programmable and abstract the underlying infrastructure for applications and network services.
The primary concept behind SDN is to enable external data control by separating the control plane from the network hardware and doing so by using a logical software entity called a controller. \cite{ordonez2017network,bouras2017sdn}. The controller is located between network devices and applications and controls packet-flow control to allow intelligent networking (Figure \ref{fig2}). Control over a network can now be programmed in this design. The 5G network can be managed by administrators and deploy new services or adjustments using the controller \cite{ordonez2017network,bouras2017sdn}.
Providers will be able to innovate, both in their operations and in their service offerings, thanks to a 5G network that is intelligent, virtualized, and programmable. They will be able to deploy new services on demand, increasing their total efficiency.
The programmability of 5G SDN networks allows new business models and accelerates revenue growth. As a result, the architecture will be flexible, cost-efficient, and dynamic, which makes it perfect for the high-bandwidth, dynamic nature of 5G use cases.

\subsubsection{Virtualization of network functions}
\begin{figure}
	\centering
	\includegraphics[scale=0.9]{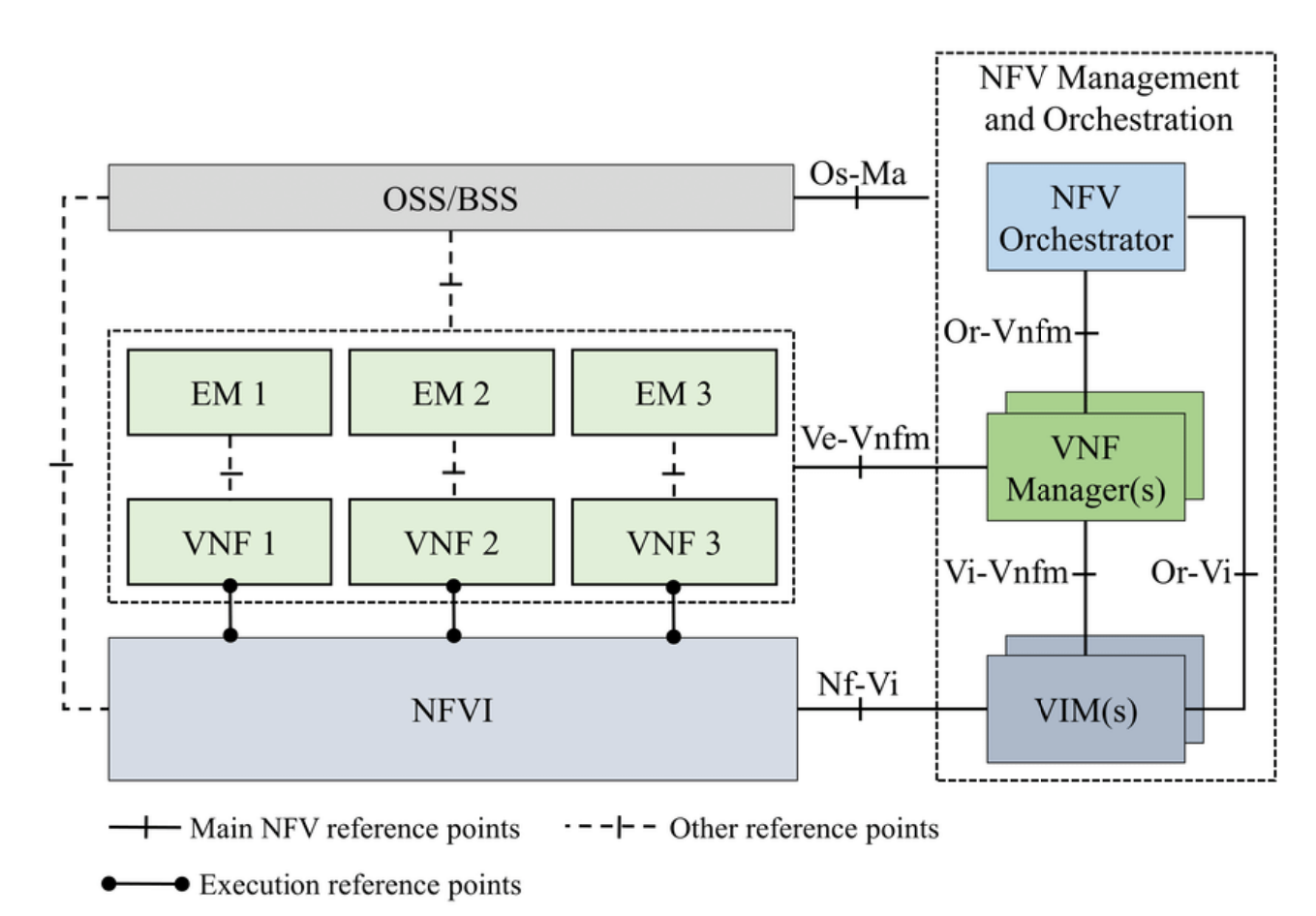}
	\caption{NFV architecture.}
	\label{fig3}
\end{figure}

Network function virtualization (NFV) is the process of substituting network functions on specialized appliances like routers and firewalls with virtualized versions running as software on commercial off-the-shelf (COTS) hardware. NFV aims to change how networks are created, and services are delivered. Each business may use NFV to simplify a wide range of network tasks, increase efficiency, and deploy new revenue-generating services more quickly and easily than ever before \cite{ordonez2017network}.
NFV is a critical enabler of the next 5G infrastructure, since it helps to virtualize all of the network's numerous equipment. NS, allows virtual network architectural component of 5G to be constructed on top of a shared physical infrastructure. Applications, devices, services, operators, and consumers may all benefit from customizing virtual networks. NFV will also permit the distributed cloud, allowing operators to build more flexible and programmable networks in the future \cite{yousaf2017nfv}. Figure \ref{fig3} illustrates the architecture of the NFV with its different entities.

\section{Network slicing}
\label{sec:others}

\begin{figure}
	\centering
	\includegraphics[scale=0.8]{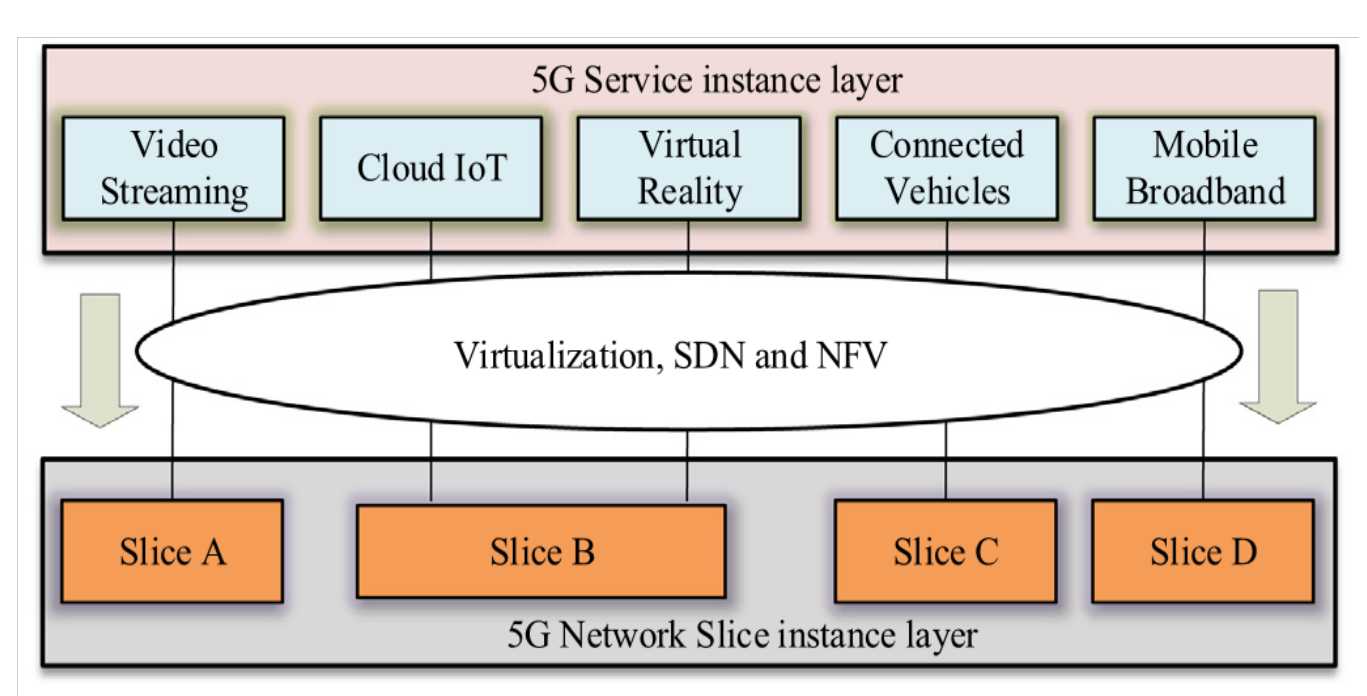}
	\caption{5G NS concept \cite{barakabitze20205g}.}
	\label{fig4}
\end{figure}

The term "5G NS" was invented and used for the first time by the next-generation mobile network (NGMN). A network slice, according to the NGMN, is an end-to-end logical network or cloud that is mutually separated and has independent management  and control plans that may be generated on demand. A network slice might include cross-domain elements coming from multiple domains in distinct or the same administrations, as well as components for the transport network, access network, edge network, and core network. Multiple logical and self-contained networks may be built on top of a shared physical infrastructure platform using the NS principle. 
NS can offer radio, cloud and networking resources to application providers or different virtual segments that have no physical network infrastructure. That way, it enables service differentiation by customizing the network operation to meet the requirements of customers based on the type of service. Figure \ref{fig4} represents the NGMN slice capabilities, where each slice represents a single or multiple service offered by a 5G network \cite{barakabitze20205g}. 

\subsection{End-to-end architecture of NS}
\label{e2eSlicing}
\begin{figure}
	\centering
	\includegraphics[]{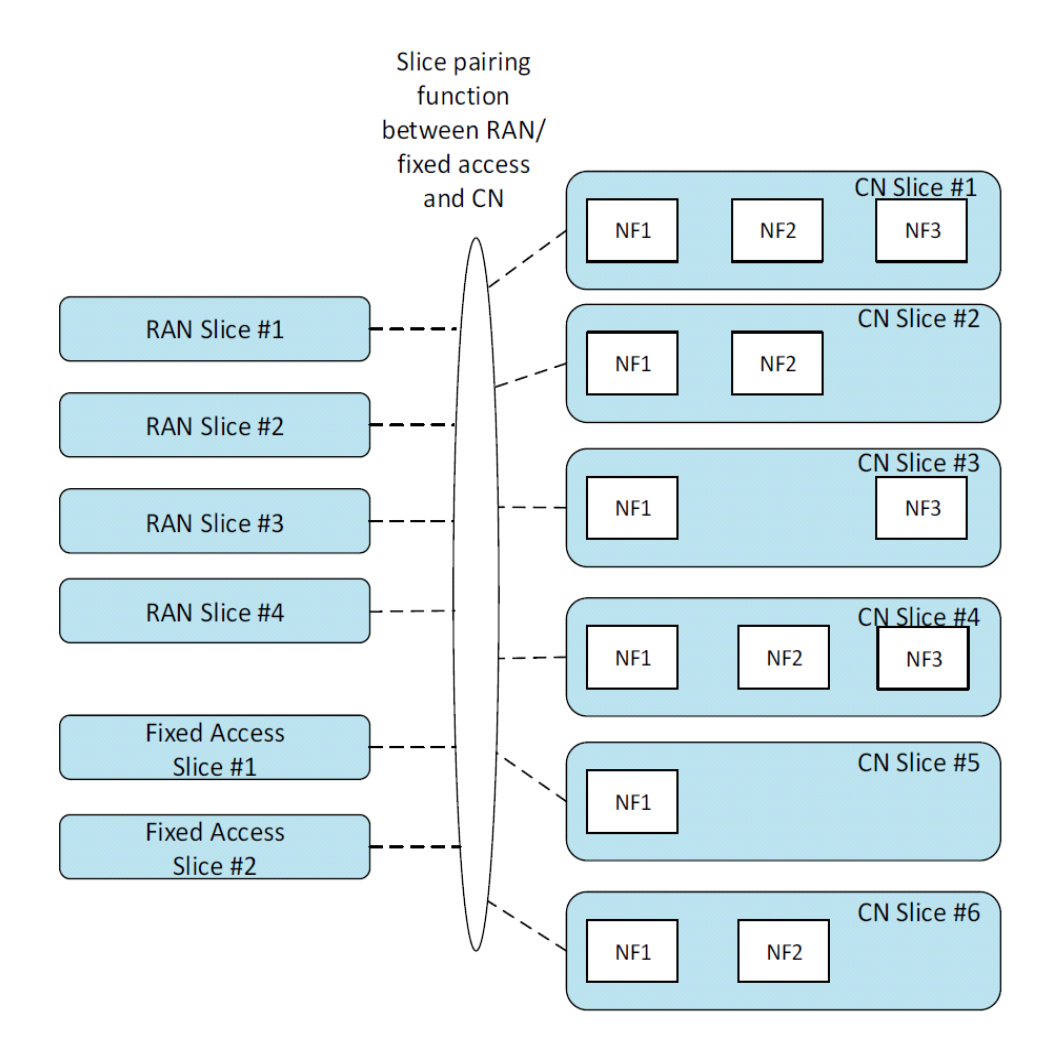}
	\caption{NS architecture.}
	\label{fig5}
\end{figure}

Access slices (ASs), 5G core network (5GCN) slices, and the selection function that joins these slices into a full network slice make up the NS architecture. The selection function directs communications to a selected 5GCN slice that provides certain services. The necessity to accommodate multiple service/application needs is one of the criteria for establishing ASs and 5GCN slices. Each 5GCN slice is made up of a collection of NFs \cite{americas2016network}. The system architecture employed in NS is shown in Figure \ref{fig5}.

\begin{figure}
	\centering
	\includegraphics[]{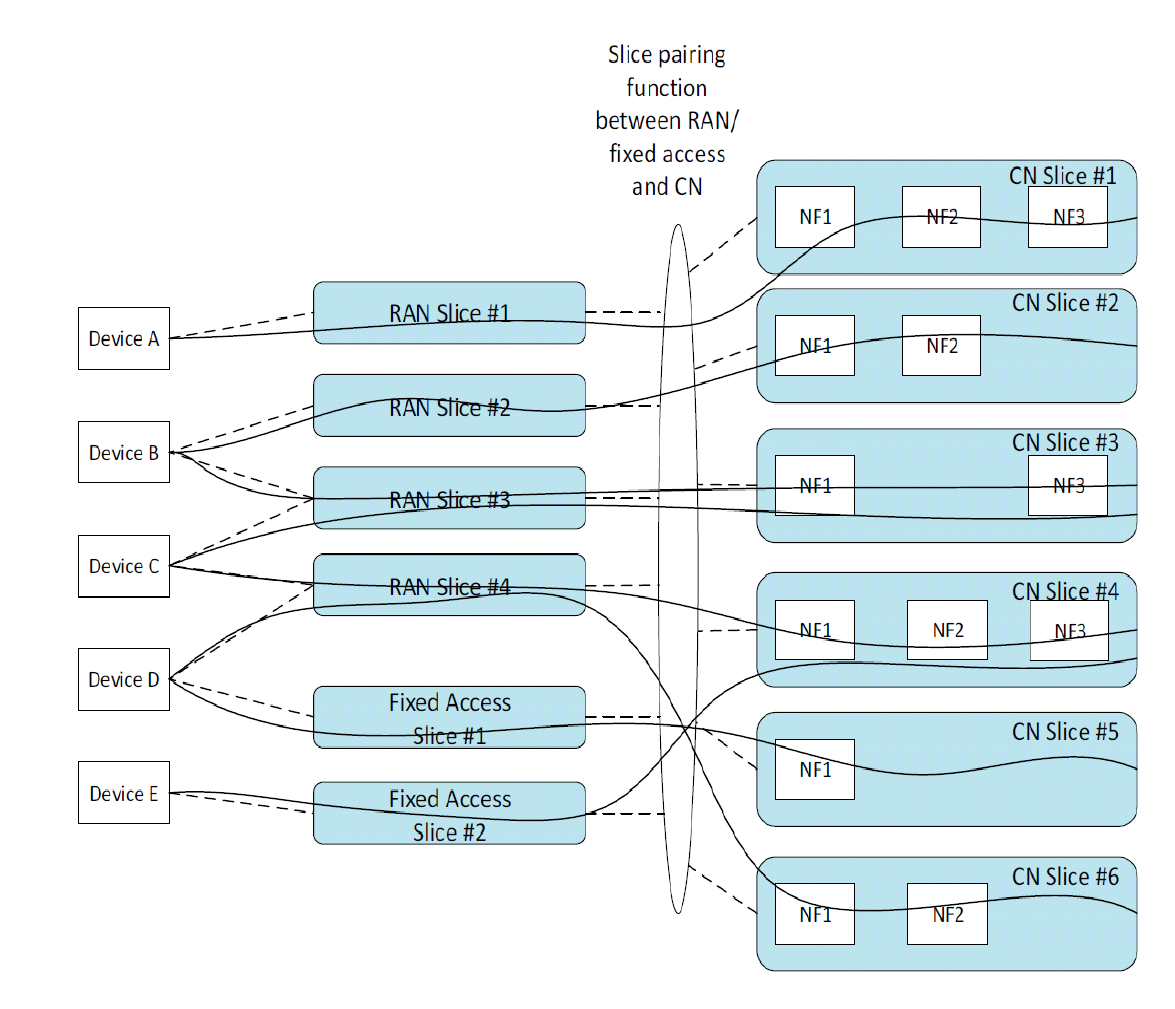}
	\caption{UE connection with the network slices.}
	\label{fig6}
\end{figure}

Some NFs may be utilized across numerous slices, while others are particular to a single slice. This is an essential consideration when slicing. The mapping between 5GCN slices, slices, and ASs can be 1:1:1 (single slice per service) or 1:M:N (multiple slices per service). Figure \ref{fig6} illustrates these two factors, for example, multiple ASs can be utilized by a single a device as device B,C and D, and multiple CN slices could connect to  an access slice like RAN slice \#3 and RAN slice \# 4.

\begin{figure}
	\centering
	\includegraphics[scale=0.9]{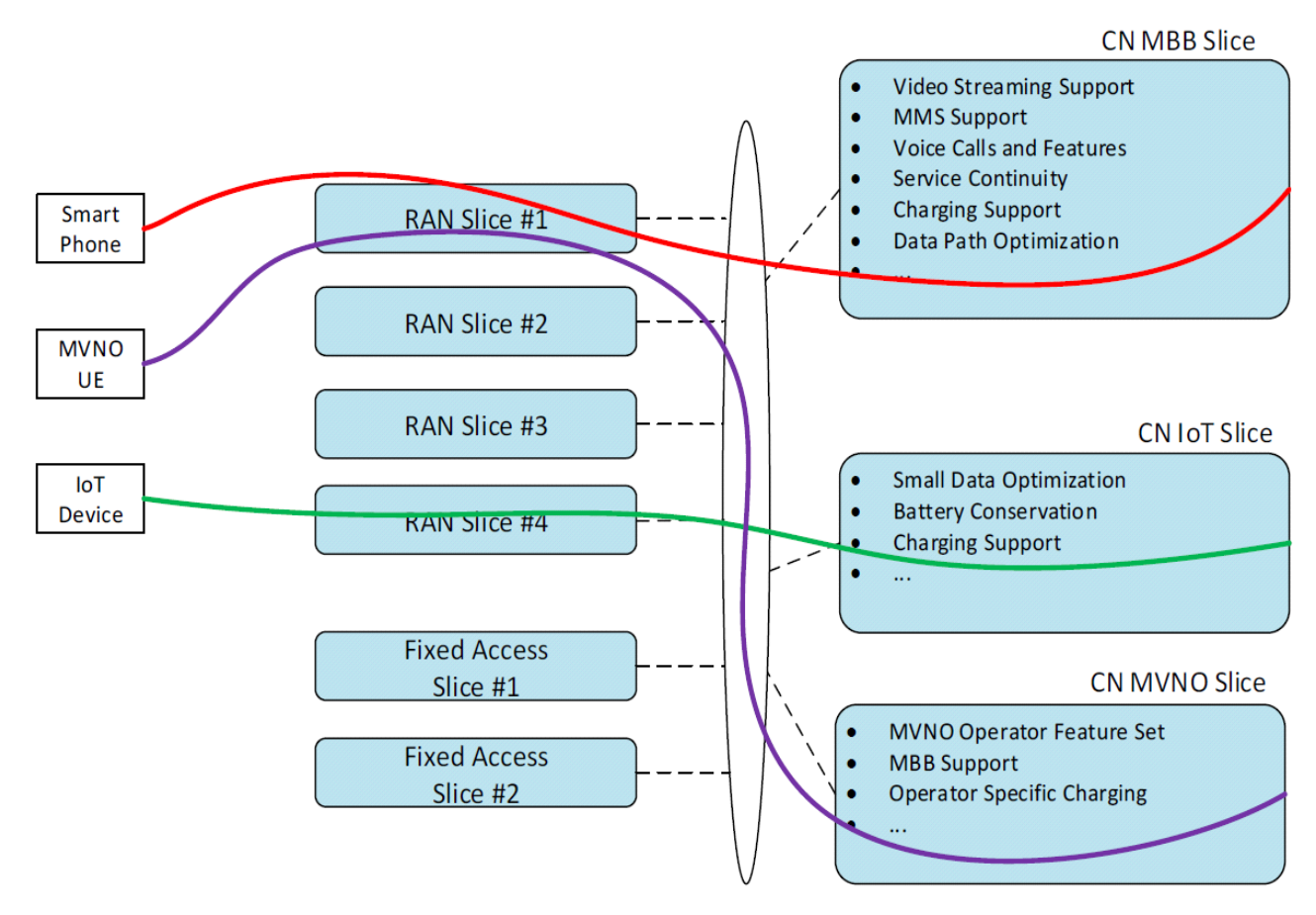}
	\caption{Network slice examples.}
	\label{fig7}
\end{figure}

To attain the required network function and communication demands, the pairing of ASs and CN slices might be static or semi-dynamic. A network slice would offer complete NF support to the devices attached to it for the duration of the targeted service lifetime \cite{americas2016network}.
Examples of network slices include: slices that remote control a factory, a virtual operator and slices that are optimized for streaming video. Figure \ref{fig7} illustrates some examples.

Some slices, such as the "Smart Phone" and "MVNO UE" devices served by RAN slice \#1, can share an access slice, as illustrated in Figure \ref{fig7}. Slices servicing various utility companies, for example, can all be linked to the access slice designed to serve mMTC, while the slice optimized for streaming video can share the mobile broadband (MBB) access slice. A dedicated access slice and a 5GCN slice would be included in some slices. A slice for factory remote control, for example, may require exclusive access and 5GCN operations to provide assured latency and dependability. A specialized 5GCN slice suited for fixed broadband access might be added to a fixed broadband access slice \cite{americas2016network}.

\subsection{Types and scenarios of NS }
In this section we mention the different scenarios that slicing can enable and the different types of NS. 

\subsubsection{	NS scenarios}

\begin{itemize}
    \item \textbf{Slicing for QoS:} Diverse applications with varying QoS needs must be served on the same infrastructure in 5G networks at the same time. However, satisfying the QoS needs of individual applications such as automobiles, smartphones, and huge internet-of-things (IoT) applications on a single network infrastructure is extremely difficult in general. Slices carry out network operations and management procedures including routing, resource scheduling, and admission control using their own unique topologies and devoted resources. Each application may be served on a distinct slice thanks to NS. A slice's setup or operation may be adjusted to meet the QoS requirement since it focuses on servicing the same kind of applications with similar QoS requirements. SDN can implement this adaptable setup \cite{an2019slice}. Slices can be allocated to individual users (Tenants) based on user demand. Various tenants might own distinct slices that service the same applications. Slices may have varying priorities depending on tenant precedence or QoS limits imposed by SLAs between tenants and infrastructure providers. Slices held by tenants that pay more for premium services, or slices with strong QoS limits on SLAs, for example, might be given more priority \cite{an2019slice}.
    
    The resources of each slice must be separated in order for the slices to provide guaranteed QoS to apps. Allocating dedicated bandwidths to the slices allows for resource separation. By controlling queues in wired networks, bandwidths may be reserved and allocated to slices. At each router, separate slice queues are conFigured and their output rates can be changed based on the requests from the slices. However, resource isolation in wireless networks may be difficult owing to radio interference across connections. A transmission of a slice interferes with other broadcasts of other slices due to the omnidirectional signal propagation property of wireless channels. This interference not only impairs the QoS of the other slices but also modifies the available real bandwidths of slices over time. As a result, it is crucial to reduce wireless interference among slices so that slices formed in a wireless network may effectively meet application QoS requirements. To minimize interference, it may be feasible to use wireless resource management technologies that assign and arrange orthogonal wireless channels among connections, such as frequency or time slot. If the number of orthogonal channels is less than the number of slices, however, interference between the slices cannot be avoided enough.
    
    \item \textbf{Slicing for infrastructure sharing:} Future 5G mobile networks will face many problems in terms of network design and essential technologies as the mobile Internet, IoT, and many new network applications and services, such as virtual reality, high-definition video, cloud services, online gaming, and so on, continue to emerge. NS has been presented as a way to deliver tailored services for various business situations in a 5G network in a flexible, efficient, and cost-effective manner. The subject of resource sharing for NS in a multi-tenant situation has been actively investigated in recent years utilizing virtualization technology, which is based on the advantages of NS in 5G networks. In-network caching, on the other hand, has been considered a promising technology in 5G networks to reduce duplicate content transmission in networks and improve end-user quality of experience (QoE) following extensive research into another new technology known as information-centric networking (ICN) \cite{jia2019caching}.
Although there have been some good publications on NS and in-network caching in 5G networks, these two critical challenges have generally been handled independently in the literature, to the best of our knowledge. Physical storage may be abstracted as a virtual cache resource pool using virtualization technologies. Furthermore, the infrastructure provider (InP) can provide cache resource slicing based on the tenant's needs. As a result, combining these two advanced strategies to increase end-user QoE and cache resource usage in 5G networks is important \cite{jia2019caching}.

The basic idea behind this NS scenario is to virtualize a wireless network's RAN domain and then distribute it across many operators. There are two types of slice tenants: slice owners and slice tenants. Based on an agreement, the owner delivers the slice to a tenant. The tenant has complete control over the slice's services and infrastructure. This notion of NS allows for the optimization of network cost models in order to increase total revenue while maintaining network scalability \cite{jia2019caching}.
\end{itemize}

\subsubsection{NS types}
Vertical and HSare two types of slicing that can be used in 5G communication networks. Vertical slicing, which focuses on 5GCN, enables vertical industries and services. HS, on the other hand, enhances system performance and improves end-user experience, and is primarily concerned with RAN infrastructure \cite{li2016end}.

\begin{figure}
	\centering
	\includegraphics[scale=0.8]{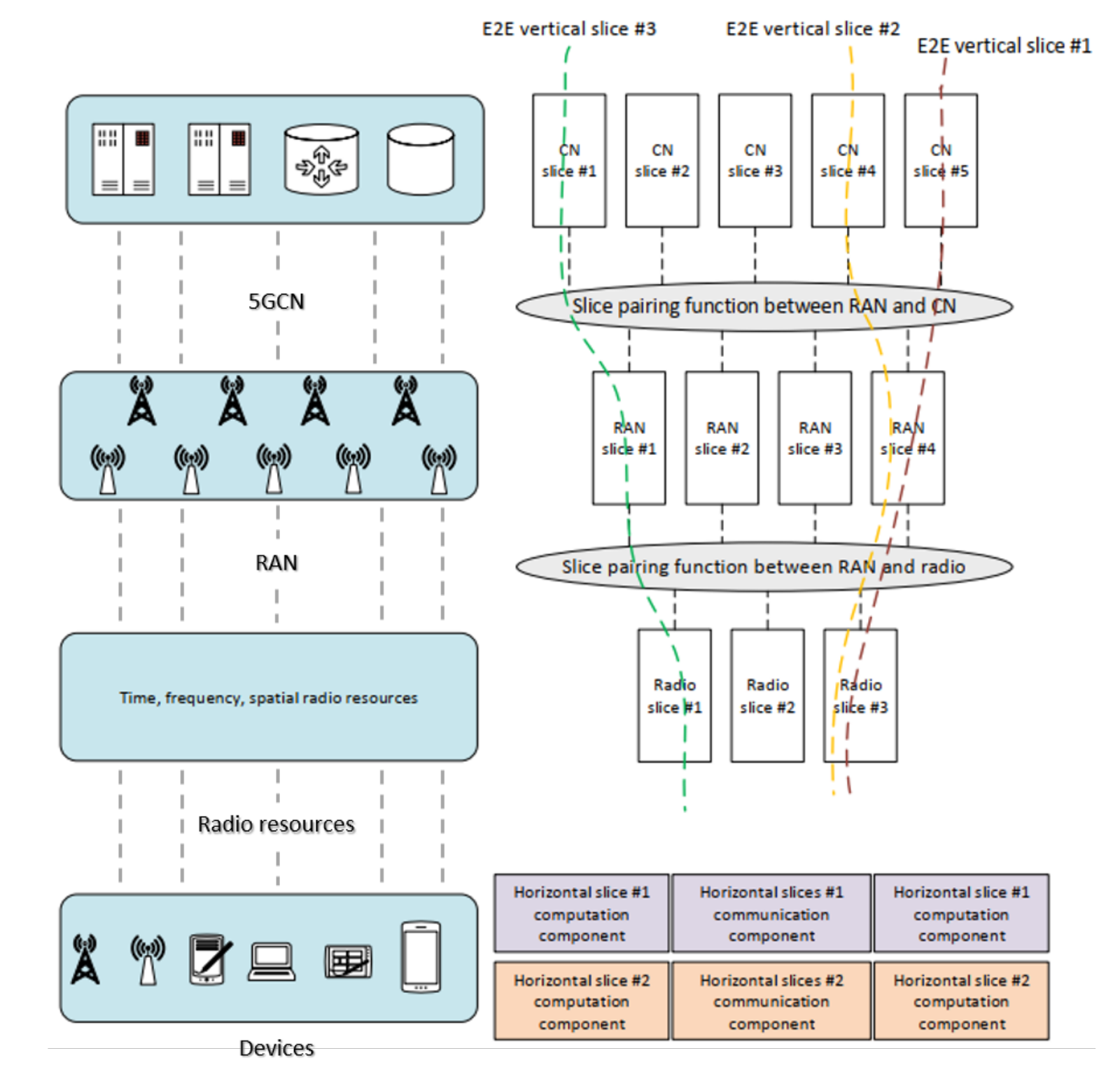}
	\caption{Illustration of vertical and horizontal NS.}
	\label{fig8}
\end{figure}

\begin{itemize}
    \item \textbf{Vertical slicing:} 
 aims to promote vertical markets and industries, and it is mostly focusing on the core domain of mobile networks. It allows applications and services to share resources and avoids or simplifies a classic QoS engineering challenge (Figure \ref{fig8}). In a vertical slice, end-to-end traffic flows between the 5GCN and the terminal devices, and each of the network nodes performs similar tasks across slices \cite{habibi2017network,li2016end}. 
 
 \item \textbf{Horizontal slicing}
is a step forward in terms of increasing mobile device capabilities and improving user experiences. HSgoes over and beyond the physical limits of platforms. HSallows a network's nodes and devices to share resources. Highly competent network devices/nodes, for example, share their resources like communication, processing, and storage with less capable network devices/nodes, resulting in improved overall network throughput \cite{habibi2017network,li2016end}. 
The notion of vertical and horizontal NS is depicted in Figure \ref{fig8}. In the vertical domain, properly designed slice pairing functions slice the physical computation, storage, or radio processing resources in the network infrastructure (as denoted by base stations and servers) and the physical radio resources (in terms of frequency, space, and time,) to form end-to-end vertical slices. When slicing the radio, the RAN, and the 5GCN, the criterion might be different. To construct end-to-end slices for various services and applications, slice pairing functions are provided to couple the radio, RAN, and CN slices. 
Physical resources (in terms of compute, storage, and radio) at nearby tiers of the network hierarchy are sliced to form horizontal slices in the horizontal domain. A machine might work on numerous slices at the same time. A smart phone, for example, can function in a vertical slice on MBB service, a vertical slice on health care service, and a horizontal slice supporting wearable devices \cite{habibi2017network}.

\end{itemize}

\subsection{NS at the RAN level}

It's necessary to be aware about two things when analyzing how NS is enabled in the RAN \cite{americas2016network}:
\begin{itemize}
    \item The slice's RAT that supports the slice's network services.
    \item The configuration of RAN resources so that they can interact with the network slice and provide support in the most effective manner feasible.
\end{itemize}
	 
The 5G infrastructure needs to meet multi-dimensional KPI goals. It is not possible to accomplish all of the goals at the same time for some of them. For example, optimizing for low latency and high reliability can often come at the cost of high spectrum efficiency, massive connection optimization, and optimizations for low latency and high reliability, as well as optimizations for high spectrum efficiency \cite{americas2016network}. 
Some conditions for the design as well as the operation of RAN slicing, which are linked with each slice to execute the network services, are:

\begin{itemize}
    \item To support each slice in the RAN, researchers have applied a list of configuration parameters to the RAN user and control plane activities. 
    \item Several slices share certain network operations (e.g., mobility management). 
    \item The use of RAN resources among the slices is coordinated by common control functions. This is done in order to improve the overall performance of the 5G network as well as its efficiency.
\end{itemize}

RAN slicing offers the opportunity to take into account the following design aspects:

\subsubsection{Management of the resource}

Radio slices in the RAN can either dynamically or statically share the radio resources (frequency, space, and time), as well as the communication hardware (analogue radio components, digital baseband processing components), depending on the configuration rules for the network slice. 

\begin{figure}
	\centering
	\includegraphics[scale=0.9]{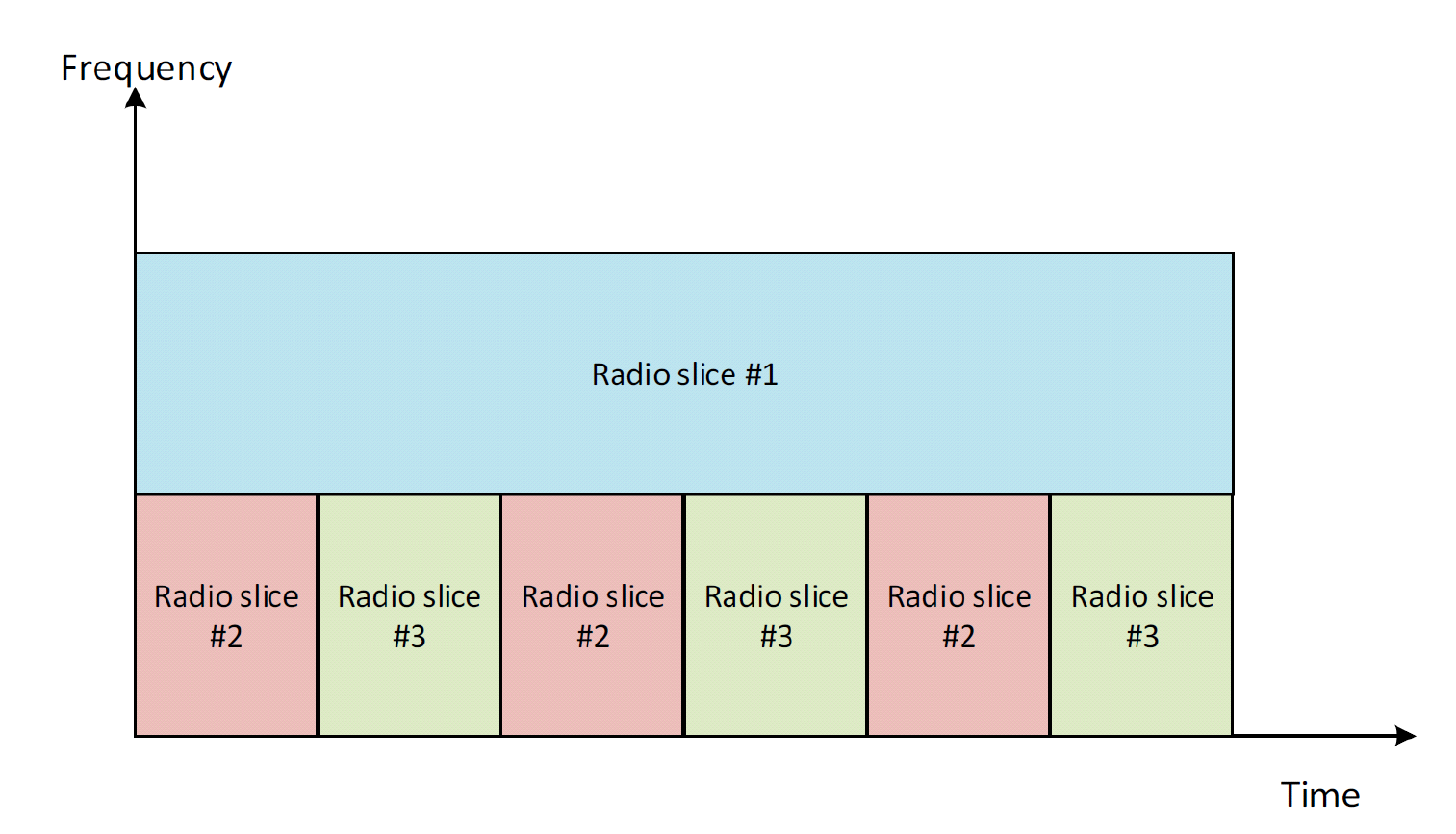}
	\caption{Multiple slices in FDM sharing a radio resource such as slice \#1 and slice \#2 or \#3, and TDM such as slice \#2 and \#3}
	\label{fig9}
\end{figure}

Each slice has access to resources depending on its priority and demand through dynamic resource sharing. Scheduling or contention can be used to share radio resources among the slices. If scheduling is used, each slice sends resource requests to a central scheduler, such as a base station scheduler or a central RAN controller. The scheduler then assigns radio resources to the slices depending on the number of resources requested, the priority of the task being performed by the slice, and the overall traffic load. In a contention-based system, each slice accesses radio resources on its own, following pre-defined criteria. A slice is pre-conFigured to run in a dedicated resource during its operating time with static resource assignment. The allocation of resources to the slices is assured with static resource sharing. Overall resource utilization can be optimized via dynamic If scheduling is used, each slice sends resource requests to a central scheduler, such as a base station scheduler or a central RAN controller. The scheduler then assigns radio resources to the slices depending on the number of resources requested, the priority of the task being performed by the slice, and the overall traffic load. In a contention-based system, each slice accesses radio resources on its own, following pre-defined criteria. A slice is pre-conFigured to run in a dedicated resource during its operating time with static resource assignment. The allocation of resources to the slices is assured with static resource sharing. Overall resource utilization can be optimized via dynamic resource sharing \cite{americas2016network}. 
Figure \ref{fig9} depicts a scenario in which each slice is assigned dedicated radio resources. FDM and TDM technologies are used to divide radio resources among the slices. Due to the fact that different radio access types have distinctively different dynamic resource configurations, the RAN configuration may vary depending on the radio access type.

\subsubsection{Control and user planes. }
In the layer 2 CP, common CP functions are defined, such as slice specific configuration rules, and device idle mode for each of the slices. It's worth noting that not all slices will require the use of all CP functions, and their RAN setting rules will specify which ones are required. In the layer 2 UP, each slice could have a specific conFigured UP protocol stack. A slice that supports several QoS classes, for example, might be set to employ distinct packet scheduling and segmentation layers (identical to LTE's MAC and RLC layers). To decrease the size of the packet header, it is possible to conFigure slices to use a single L2 UP if they send small packets and have just one QoS class. As an element of a hierarchical MAC layer, it is possible to specify either an outer layer MAC for intra-slice scheduling or an inner layer MAC for inter-slice scheduling \cite{americas2016network}.

\subsubsection{Control of specific slice admission}

To fulfill the initial access needs of distinct network slices, admission control is required. The capacity to accommodate the access demands of a range of network slices is provided by RAN configuration rules. A network slice that serves mission-critical services, for example, must have assured low-latency access. A UE operating in one slice may be denied admittance to an access point if that slice is not active in that access point, according to slice-specific admission control configuration rules. The common admission control has the ability to activate slices in the RAN that either have not been activated at an access point or slices that do not need a slice-specific admission control in order to function properly. \cite{americas2016network}. 

\subsubsection{UE comprehension of RAN configurations}

The RAN settings for the various services may or may not be known to the UE. The UE seeks access to a service specified by one core network slice in the first case, and the RAN assigns appropriate radio resources to provide the service \cite{barakabitze20205g}. The UE might be unaware of the RAN resource configuration. In the second situation, the UE requires defined RAN configuration and explicit system information signaling before being able to access the service, such as non-default services provided by the RAN. 

\subsection{NS management framework }

A network slice model is defined by NGMN as a detailed description of the configuration, work flows, and structure for how to construct and govern a network slice instance throughout its life cycle. The architecture for end-to-end slicing, presented in section \ref{e2eSlicing}, is a logical decomposition of the network slice instance layer that considers particular network domain functionalities like 5GCN and radio network domains \cite{americas2016network}.
From an operational standpoint, network operators must meet the following fundamental conditions for NS activities, as defined by the 3GPP:
\begin{itemize}
    \item Create and control services for the pairing of slices feature.
    \item Create blueprints for network slices.
    \item Deploying and managing fault, configuration, accounting, performance, and security features for the specified network slices is necessary. Per slice, performance management (PM), fault management (FM), and configuration management (CM) should be available. As a result, operators may track end-to-end service that includes important data on the performance of slice in relation to the needed QoS as well as the performance of the NFs.
    \item Create and maintain the pairing function for slice components (e.g., CN slice to RAN slice pairing rules).
    \item Network slices may be modified, including adding, removing, and altering network slices.
    \item Because the management of resources for a network slice might cross many operator domains, it is possible that multiple resource management domains cooperating together is necessary.
    \item The slice-as-a-service idea enables a third party to design and maintain network slice configurations (e.g., scale slices) using appropriate application programming interfaces (APIs) while staying within the operator's constraints.
\end{itemize}

\begin{figure}
	\centering
	\includegraphics[scale=0.9]{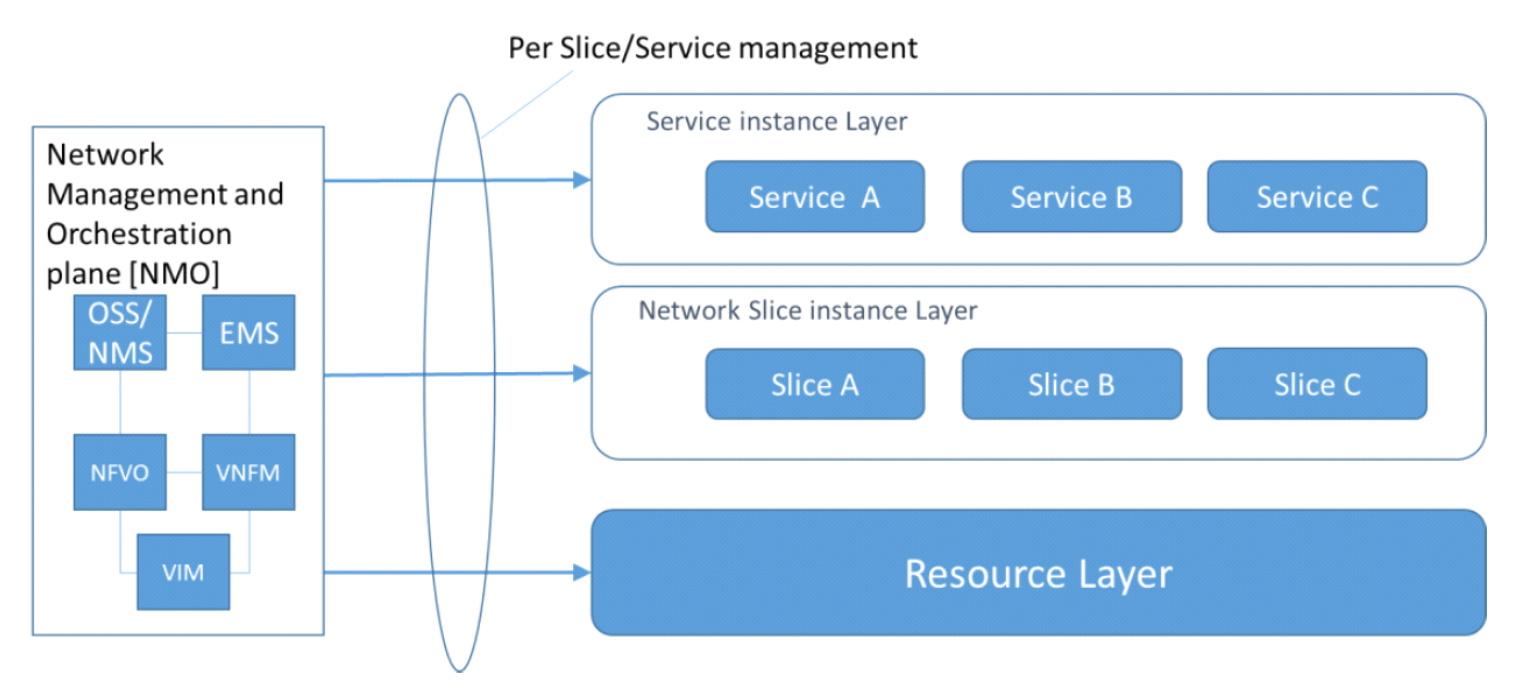}
	\caption{Network slice management.}
	\label{fig10}
\end{figure}

The network management and orchestration (NMO) plane, seen in Figure \ref{fig10}, offers management and orchestration services for the three levels that make up the sliced network architecture. Each slice is orchestrated and managed individually by NMO functions. 

\begin{figure}
	\centering
	\includegraphics[scale=0.8]{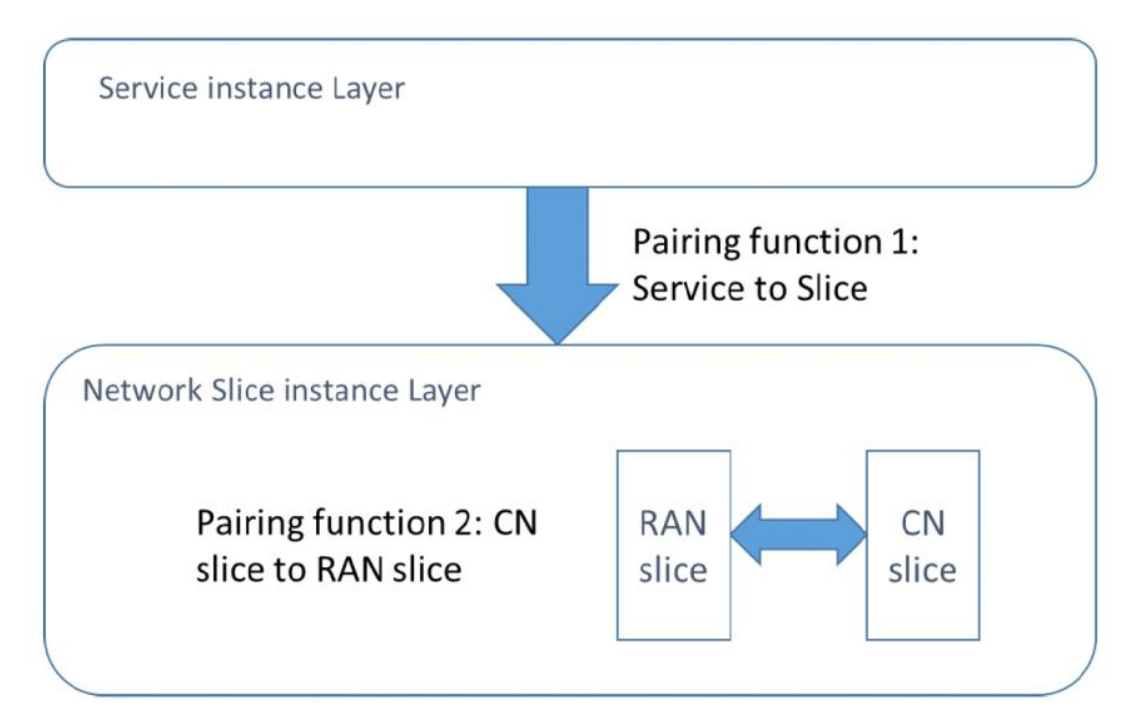}
	\caption{Service to network slice pairing.}
	\label{fig11}
\end{figure}

As mentioned in Figure \ref{fig10} , the management of NS has to take into consideration the appropriate mapping functions that are utilized to construct a slice. Besides, it must consider the life cycle decisions and the service provider's criteria to establish the mapping's initial state depending on network and use dynamics \cite{americas2016network}. The necessary pairing functions that must be maintained for NS are shown in Figure \ref{fig11}.

\begin{figure}
	\centering
	\includegraphics[scale=0.8]{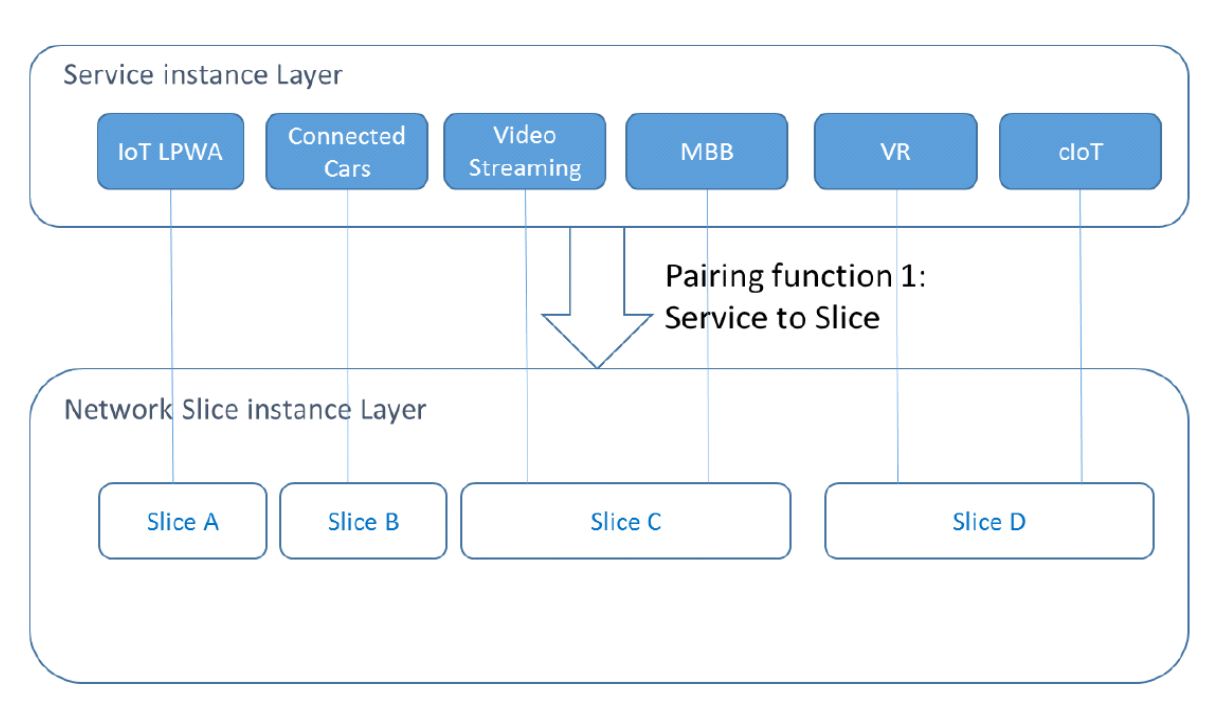}
	\caption{Service instance to network slice instance.}
	\label{fig12}
\end{figure}

Figure \ref{fig12} depicts how different slices have the possibility to map services in accordance with business strategies and the policies of operators. Each slice has a set of characteristics, such as end-to-end low latency or high bandwidth, that make it more suitable for specific types of applications.

From an operational standpoint, the management system must enable operators to map services to slices, such as: 

\begin{itemize}
    \item A provided service must be mapped to a specific network slice by the operator.
    \item A network slice is a collection of network capabilities and SLAs that are suited for a variety of service kinds. As a result, a single network slice may be mapped to several services. An eMBB-oriented slice, for example, could be able to support music downloads, video streaming, and huge file transfers. Another low-latency slice, meanwhile, may handle real-time communications and VoIP.
\end{itemize}

\begin{figure}
	\centering
	\includegraphics[scale=0.9]{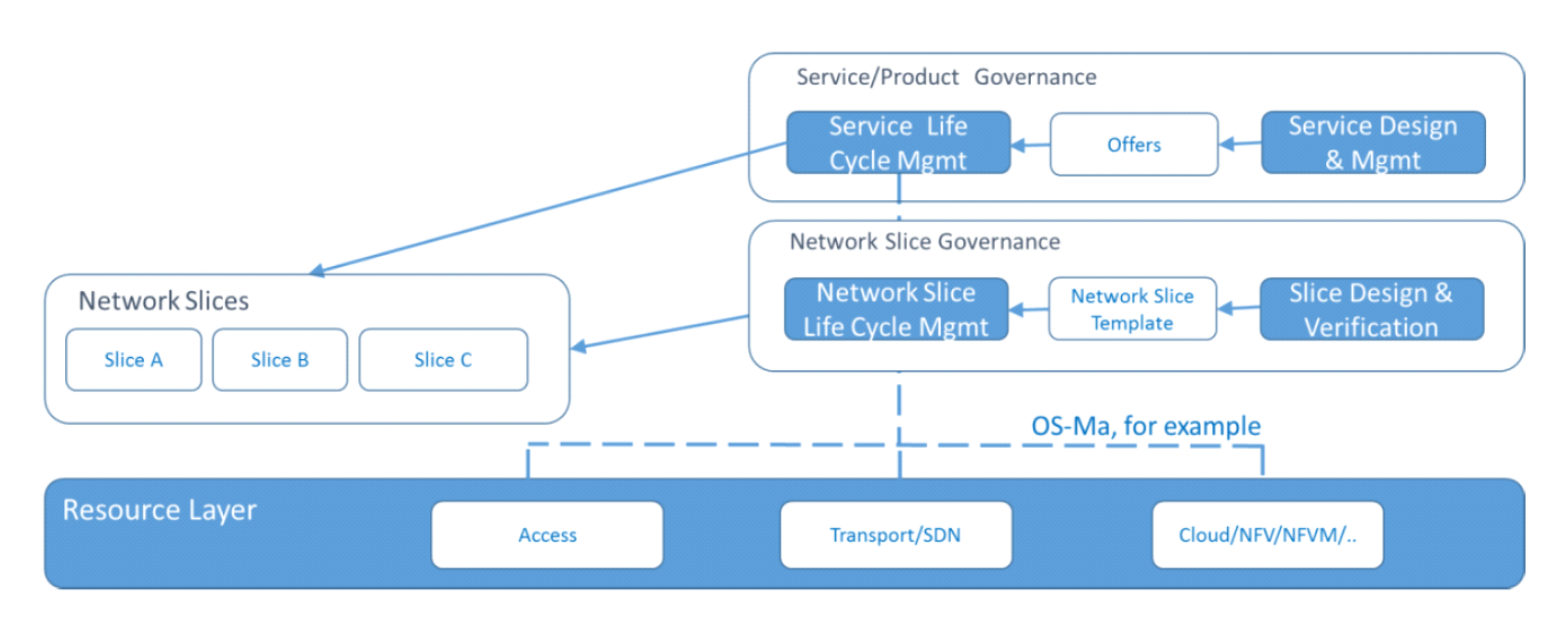}
	\caption{Management and resources for network slices.}
	\label{fig13}
\end{figure}

Different slices may be assigned to users that have the same service type. When tenant isolation is required, every slice is allocated to a certain group of clients such as local and roaming users. One such example is the use of geographical distance, which directs users to the slice nearest to them in order to decrease backhaul expenses and delay. A network slice might be allocated to a certain type of service. For example IoT, communications may be directed to a slice with minimal latency and features, which is optimized for IoT devices. Multiple administrative domains may be involved because of a network slice nature. This includes virtualized and non-virtualized wireless and wired   resources of the network. Roaming is a circumstance in which a network slice instance travels through many administrative domains. Support for business verticals is another example, where different administrative domains may be able to give extra capabilities. Services, network slice instances, and network service instances must thus be managed across various management domains. The service will require a NS orchestration function to manage NS. The resource orchestrator, which is made up of network functions virtualization orchestration (NFVO) and 3GPP application resource configurators, is responsible for network slice management at the resource layer \cite{americas2016network}.
In Figure \ref{fig13}, the OS-Ma interface is used to demonstrate slicing governance.

\subsection{Opportunities introduced by the NS}

Operators may now differentiate themselves through service personalization thanks to the advent of slicing. Various automobile sellers, for example, may operate in different slices when it comes to linked cars. Operators will be able to assign specialized virtual resources to various automobile manufacturers and run them independently. Different services, such as traffic efficiency and safety, news as well as entertainment, urgent assistance help and security, may be available from different automobile manufacturers. Mobile carriers can offer distinct pricing strategies by differentiating their services. Another area where NS might help operators is in monetizing QoE. In 5G, QoS may be defined based on the specific requirements of each application.  An operator may construct several slices for various VNOs, each with a distinct QoE value, allowing them to monetize their services.
One of the most significant difficulties that today's operators face is that the network may get congested with a few malfunctioning devices, reducing the existence and dependability of network resources for other lawful devices. NS gives operators another option in which particular types of traffic are confined within a slice and the behavior of that network segment has no effect on the other slices. A faulty sensor or actuator in one network slice will have no effect on in another key service, such as public safety.
NS might be a best way to resolve one of the wireless industry's problems: how to increase market growth to serve sectors and vertical services. The wireless sector must keep enhancing user experience, expand equipment capabilities, as well as handle traffic scalability in the future \cite{americas2016network}. Slicing technology can be used to overcome these issues. Through the air interface, computing resources from network infrastructure and mobile devices may be sliced and combined to construct virtual computation platforms. Communication and computing work together to meet capacity scalability and device capability augmentation goals, thanks to the virtual computation platform. Techniques like edge computation, for example, can terminate traffic at the network edge, easing capacity scaling demands in the deeper network architecture. 

Densification and radio resource reuse at the network edge provide higher capacity growth \cite{barakabitze20205g}. Additionally, communication enables the offloading of computing from low-capability devices, such as wearables, to high-capability devices, such portable devices or network infrastructure, allowing the low-capable potential of device to be augmented beyond its physical limitations.

\section{An example of NS }

From the perspective of operators, the evolution of mobile networks is a never-ending quest to find a balance between capacity expansion and cost reduction. NFV enables providing customized services to users, due to the fact that any virtual network operator (VNO) may create their own protocols over the shared radio resources. Wireless network virtualization, on the other hand, allows for the deployment of convergent resource management mechanisms, which is critical for heterogeneous networks that are always changing (Het-nets) \cite{caeiro2017fair,rouzbehani2018optimised}.
In order to offer each VNO with the functionality they want, numerous VNOs must coordinate their services via orchestration, it is one of the major challenges in NVF. Each service must be managed in isolation because vertical applications have diverse and often orthogonal service requirements. To cater application-specific customization while ensuring an effective way of using radio resources across slices, convergence among VNOs must be achieved. Those having a network-wide approach for radio access slicing are more effective ways for RRM specifically in Het-Nets. In the mobile network architecture, such models often propose a logical management entity on top of base stations \cite{caeiro2017fair,rouzbehani2018optimised}. RRM models that reach entire network not only allow dynamic resource sharing, but they also need minimum adjustments to conventional base stations.
In this part, an RRM technique that is applicable across the network's virtual RANs is provided. The primary contribution is the suggestion of a model that can both fulfill the concept of service-oriented separation between the responsibilities of InPs and VNOs and also support multi-tenancy. \cite{caeiro2017fair,rouzbehani2018optimised}. In this case, various parties request Capacity-as-a-Service (CaaS) from a centralized virtualization platform in order to meet various contractual SLAs. The fundamental purpose is to maximize the utilization of the entire available capacity given by the aggregation of radio resources from multiple RATs by serving users according to a proportional fairness framework with pre-defined serving weights and priority .

\subsection{Network architecture of the model }

\begin{figure}
	\centering
	\includegraphics[scale=0.8]{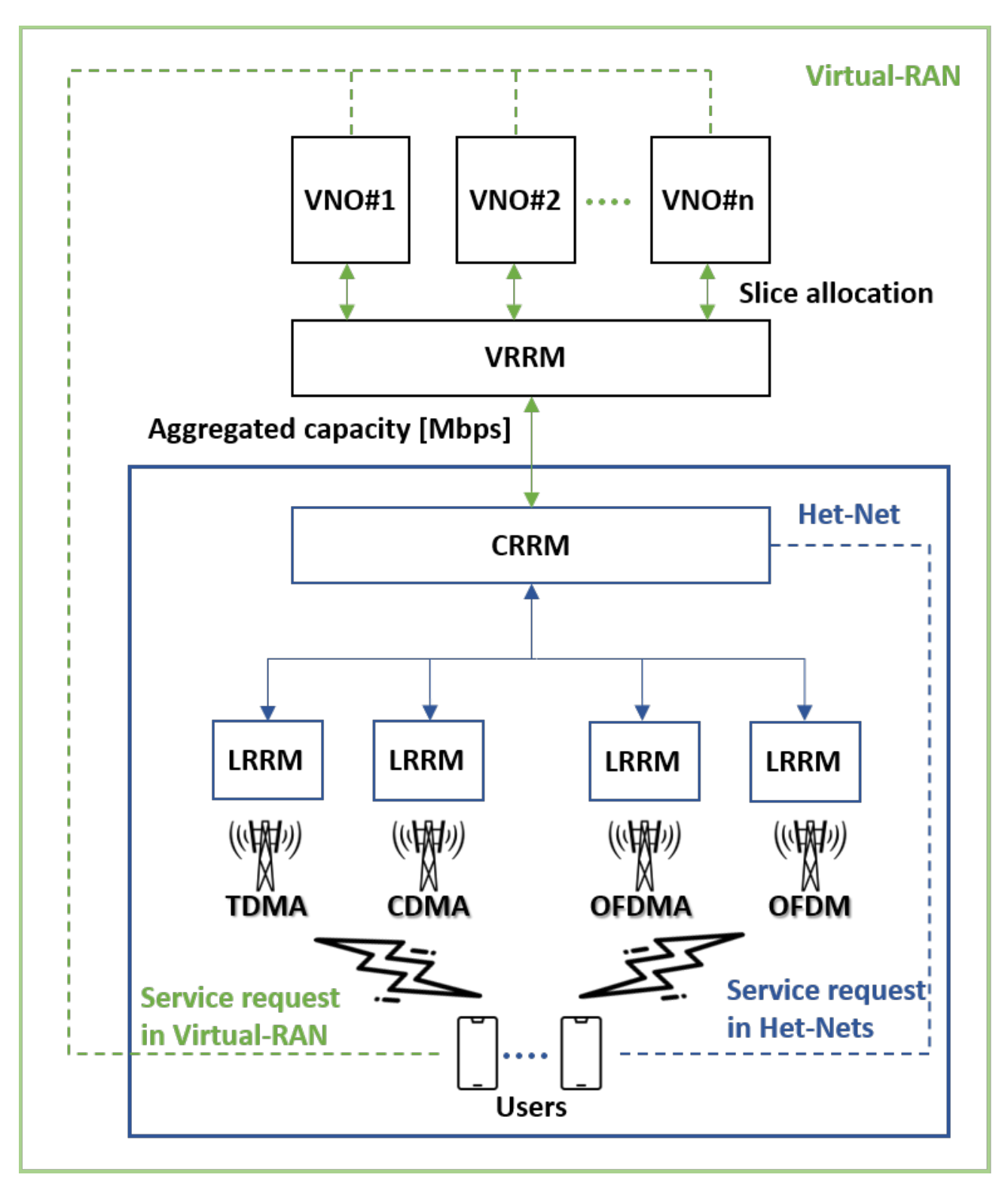}
	\caption{Network architecture of the virtual RAN \cite{caeiro2017fair}.}
	\label{fig14}
\end{figure}

In this part the conceptual architecture of the network with the functionalities of the components involved. In order to highlight the main difference between the performance of traditional Het-Nets and the model proposed for virtual RANs and to explain the functionalities of the parties involved in the model, Figure \ref{fig14} proposes a hierarchical network architecture based on softwarization, which is a conceptual architecture consistent with the business model suggested by 3GPP, in which InPs do not offer services to end users. Each VNO requests capacity as a service from the centralized virtualization platform virtual radio resource management (VRRM). The common radio resource management (CRRM) is responsible for managing the overall available capacity, and the VRRM controls this capacity by aggregating all radio resource units (RRUs) from various RATs. These RRUs can be OFDM (carriers for Wi-Fi), OFDMA (resource-blocks for LTE), CDMA (codes for UMTS) and TDMA (time-slots for GSM). The VRRM allows each VNO to deploy its own protocol stack on the same type of RRUs per RAT. The required capacity must be delivered in compliance with the SLA agreements between the VNOs and the InP \cite{caeiro2017fair,rouzbehani2018optimised}.
The low-level physical RRUs are aggregated, abstracted, and virtualized in NS in order to offer high-level resources. This makes it possible for numerous VNOs to share a base station (BS) or access point (AP). The deployment of distinct protocol stacks is made possible by NS, which enables VNOs to share radio resources. As a direct consequence of this, the usable radio spectrum is considered a single resource that has to be partitioned into virtual resource slices. In addition, the procedure for resource allocation may be summed up as follows: users make service requests to the VNO that is linked to CRRM, and the VNO tells VRRM about the total capacity that is available based on the information provided by the users. The necessary data rates are distributed between the services using VRRM. 
Three VNOs are proposed and which correspond mainly to the 3 major areas of application
of the next fifth-generation (5G) mobile telephone network, in particular eMBB, mMTC, and  uRLLC. These three aforementioned VNOs represent three (03) different categories of SLA service contracts depending on the data rate perceived by the user (User Data Rate), distributed as follows \cite{caeiro2017fair,khatibi2017modelling}:

\begin{itemize}
    \item Guaranteed bitrate (GB), in which the RAN provider promises the VNO a minimum and maximum data rate despite of network state. The VNO is completely satisfied when the maximum promised data rate is assigned to it. VNOs can have complete control over their networks thanks to the upper barrier in this sort of SLA.
    \item 	Best effort with minimum guaranteed (BG), where a minimum level of service is guaranteed to the VNO. Requests for data rates greater than the specified level are handled with the utmost care; as a result, the minimum guaranteed data rate is the one obtained during peak hours. In this circumstance, although not investing as much as in the past, VNOs may nevertheless guarantee a minimal level of QoS to their subscribers. From the perspective of the subscribers, the adequate service (but not as good as the former ones) is provided at a reduced cost.
    \item 	Best effort (BE), in which the VNO is delivered with the pure best effort approach. Operators and consumers may experience reduced QoS and resource scarcity during peak hours in this instance, but the accompanying cost will be lower as well.
\end{itemize}
	
\subsection{Model objective}
The main objectives of this model are:

\begin{itemize}
    \item Promote the coexistence of a multiple VNO of the same InP.
    \item A distinction should be made between the functions of the InPs and the VNOs.
    \item Assurance of the principle of CaaS to satisfy the different SLAs.
    \item It is important to get the most out of the entire capacity that is supplied by the aggregation of resources.
\end{itemize}

This model is initially based on the following two concepts:

\subsubsection{	Resource allocation}

When allocating resources in Het-Nets, the VRRM model is designed to strike a balance between efficiency and fairness. Sm detailed in Equation 1, Maps a part of network bandwidth that users employ it a measure of efficiency, quantifying the presumed satisfaction of users given in the allocated resources \cite{caeiro2017fair}. 

\begin{equation}
    sm=\max_{w^{usr}}\sum_{vs=1}^{N^{srv}} \lambda_{vs} log \Bigg( \sum_{i=1}^{N_{vs}^{usr}} w_{vs,i}^{usr} \frac{R_{vs}^{srvMax}}{R^{CRRM}} \Bigg)
\end{equation}

Where, \\
$w^{usr}$: users weights vectors.\\
$\lambda_{vs}$: tuning weight of service s from VNOs.\\
$N_{vs}^{usr}$: number of users per service s.\\
$N^{srv}$: total number of services of all VNOs combined.\\
$R^{CRRM}$: total offered capacity from CRRM to VRRM.\\
$R_{vs}^{srvMax}$: maximum data rate that can be assigned to a VNO service user. \\
Such that:

\begin{equation}
    \lambda_{vs}=\gamma_v.\delta_s
\end{equation}
where, \\
$\delta_s$ refers to service s weight from VNO. \\
$\gamma_v$ refers to VNO weight.\\

By optimizing the allocation of radio resources for different applications, VRRM creates a bridge between the functionalities of the MAC and higher layers.
It is useful to point out that $\delta_s$   projects the internal policy of each VNO in the context of capacity distribution through the services provided and that its values are fixed for each VNO in order to facilitate the calculations. This nonlinear optimization function will make it possible to find the optimal values of the vector relating to the weight of users in order to satisfy the various customers of the SLAs available, subject to the linear constraints mentioned in section  \ref{constraints}.
The allocation principle is based on priority based on weight, for example, a service with a higher priority receives a higher service weight and vice versa. The notion of weight assignment is used to determine the order of priority for allocating data rates to the various services provided by the various VNOs. Services that have a greater weight are provided at higher data rates. The level of service contracts that exist between VNO and VRRM served as the basis for the selection of these weights \cite{caeiro2017fair}.

\subsubsection{The constraints}
\label{constraints}
There are constraints in the allocation of data rates that must not be violated and must be respected in order to achieve the aforementioned objectives while satisfying the optimization function put in place \cite{caeiro2017fair}.

\begin{itemize}
    \item \textbf{Constraint with respect to the data rate of the service s:} The second set of limitations that must be met is imposed by the degree of service that must be assured. The data rates that are allocated to each service should be higher than the minimum that is promised. There is no upper limit for best effort services or best effort services with guaranteed minimum services; nevertheless, the data rate assigned for guaranteed services is restricted to the utmost that can be provided. VRRM's goal function must be solved while maintaining to certain limitations. Each user's average long-term data rate must fall within the VNOs-defined tailored range of data rate changes, which may be represented as detailed in Equation \ref{eq3}.
    
    \begin{equation}
    \label{eq3}
        R_{vs}^{srvMin} \leq w_{vs,i}^{usr}. R_{vs}^{srvMax} \leq R_{vs}^{srvMax}
    \end{equation}
where, $R_{vs}^{srvMin}$ refers to minimum data rate that can be assigned to a user of service s, according to VNO.   
\item \textbf{ Constraint against the total capacity allocated to each VNO:} Further constraint is the capacity between VNOs and InPs, which varies depending on the SLA type This latter can be expressed as the total throughput of all users using the different services of any VNO, in other word, it represents a range of data rates acceptable for each VNO varied between min and max. Formally, the total capacity constraint allocated in each VNO is expressed in Equation \ref{eq4}:

\begin{equation}
    \label{eq4}
        R_{vs}^{vnoMin} \leq \sum_{vs=1}^{N_{v}^{srv}}\sum_{i=1}^{N_{vs}^{usr}}  w_{vs,i}^{usr}.  R_{vs}^{srvMax} \leq R_{vs}^{vnoMax}
\end{equation}

Where:
$R_{vs}^{vnoMax}$ and $R_{vs}^{vnoMin} $ refer to the max, the min contracted data capacity of the VNO with the InP.

$N_{v}^{srv}$ refers to the number of services provided by the VNO.

\item \textbf{ Constraint in relation to the total aggregate maximum capacity of the network: } 
The principle constraint is respecting the total network capacity provided by the CRRM. The sum of all data rates assigned to all services of all active VNOs must not exceed the total aggregate capacity of the network, and can be calculated according to Equation \ref{eq5}.

\begin{equation}
    \label{eq5}
        \sum_{vs=1}^{N_{v}^{srv}}\sum_{i=1}^{N_{vs}^{usr}}  w_{vs,i}^{usr}.  R_{vs}^{srvMax} \leq R^{CRRM}
\end{equation}
\end{itemize}

The proposed problems could be solved by CVX optimization tool \cite{grant2014cvx}.

\section{CVX optimization tool}

CVX is a MATLAB-based convex optimization modeling system. It converts MATLAB into a modeling language, allowing users to specify restrictions and objectives using regular MATLAB expression syntax. CVX enables a disciplined convex programming approach to convex optimization in its default mode. Convex functions and sets are constructed with this method using a minimal number of principles from convex analysis and a base library of convex functions and sets. These rules automatically translate constraints and objectives into a canonical form and solve them \cite{grant2014cvx}. 

Consider the example of convex optimization model below: 

\begin{align*}
    \text{Maximize } \| Ax-b \|_2 \\
    \text{Subject to }  Cx=d \\
    \|x\|_\infty \leq  e 
\end{align*}

The line of MATLAB code that follows can be used to produce and solve a random instance of this model:

\begin{lstlisting}
m = 20; n = 10; p = 4; % Initialization of the variables 
A = randn(m,n); b = randn(m,1);
C = randn(p,n); d = randn(p,1); e = rand;
cvx_begin % creates a placeholder for the new CVX specification
    variable x(n) % Declares x to be an optimization variable of dimension n
    minimize (norm(A * x - b, 2)) % specifies the objective function to be
    minimized
    subject to % Adding the constraints
    C * x == d 
    norm(x, Inf)<= e
cvx_end 
\end{lstlisting}

The least-squares problem is solved when MATLAB reaches the cvx\_end command, and the result is written to MATLAB variable \cite{grant2014cvx}.

\section{Case scenario study}

In this scenario, there are three VNOs, each with a different SLA contract (GB, BG, BE). Table \ref{tab2} shows the different parameters and priorities of the three VNOs used for this study.

\begin{table}[H]
\centering
\caption{VNO’s parameters and priorities.}
\label{tab2}
\begin{tabular}{lllll}
\hline
VNO & SLA & Weight & $R^{vnoMax}$ & $R^{vnoMin}$ \\
\hline
1 & GB & 10 & 0.7 $R_{CRRM}$ & 0.4 $R_{CRRM}$ \\
2 & BG & 5 & $R_{CRRM}$ & 0.4 $R_{CRRM}$ \\
3 & BE & 1 & $R_{CRRM}$ & 0 \\
\hline
\end{tabular}
\end{table}

The network parameters specifications are presented below (Table \ref{tab3}). Each one of the three VNOs is in charge of providing three different types of service: voice, IoT and eMBB. As well as the user mix (\% share) of each service, and their allocated weights as mentioned above. It also shows the different data rates based on the deferent SLAs contracts. Lastly the maximum data rate capacity of the totality of the network is set to 630Mbps ($R_{CRRM}$).

\begin{table}[H]
\centering
\caption{Network parameters}
\label{tab3}
\begin{tabular}{lllllllll}
\hline
$R_{CRRM}$ & VNO & Service & $R^{srvMin}$  & $R^{srvMax}$ & User \% & $R^{vnoMin}$  & $R^{vnoMax}$ &  \\
\hline
 &  & Voice & 0.032 &  0.064 & 20 &  &  &  \\
 &  & IoT & 0.5 & 1 & 50 &  &  &  \\
 & \multirow{-3}{*}{GB} &  eMBB & 4 & $R_{CRRM}$& 30 & \multirow{-3}{*}{\begin{tabular}[c]{@{}l@{}}0.4 $R_{CRRM}$\\    \\ =252\end{tabular}} & \multirow{-3}{*}{\begin{tabular}[c]{@{}l@{}}0.7 $R_{CRRM}$\\    \\ =441\end{tabular}} &  \\
  \cline{2-8}  
 & & Voice & 0.032 &  0.064 &  20 &   &   &  \\
  &   &  IoT &  0.5 &  1 &  50 &   &   &  \\
  & \multirow{-3}{*}{ BG} &  eMBB &  4 &  $R_{CRRM}$ &  30 & \multirow{-3}{*}{ \begin{tabular}[c]{@{}l@{}}0.4 $R_{CRRM}$\\    \\ =252\end{tabular}} & \multirow{-3}{*}{ \begin{tabular}[c]{@{}l@{}}$R_{CRRM}$\\    \\ =630\end{tabular}} &  \\
   \cline{2-8}  
  &   &  Voice &  0 &  0.064 &  20 &   &   &  \\
  &   &  IoT &  0 &  1 &  50 &   &   &  \\
\multirow{-9}{*}{ 630} & \multirow{-3}{*}{ BE} &  eMBB &  0 &  $R_{CRRM}$ &  30 & \multirow{-3}{*}{ 0} & \multirow{-3}{*}{ \begin{tabular}[c]{@{}l@{}}$R_{CRRM}$\\    \\ =630\end{tabular}} & \\
\hline
\end{tabular}
\end{table}

With the help of CVX implemented in MATLAB, a simulation was ran to dictate the weights and data rates allocated to individual users of each of the 12 services. The preliminary results obtained from $w_{vs,i}^{usr}$ with the data rates provided and corresponding to each service are given in Table \ref{tab4}.

\begin{table}[H]
\centering
\caption{Results obtained from $w_{vs,i}^{usr}$ with service flow rates s.}
\label{tab4}
\begin{tabular}{llllllllll}
\hline
  $R_{CRRM}$ &   VNO & Service & $R^{srvMin}$  & $R^{srvMax}$  & \begin{tabular}[c]{@{}l@{}}Service \\ weight\end{tabular} & \begin{tabular}[c]{@{}l@{}}VNO \\ weight\end{tabular} & $w^{usr}$ & \begin{tabular}[c]{@{}l@{}}$R^{srv}$ \\ provided \end{tabular}& \begin{tabular}[c]{@{}l@{}}$R^{vno}$\\ provided \end{tabular}  \\ 
\hline   
   &    & Voice & 0.032	 & 0.064 & 5 &    & 1 & 0.064 &    \\
   
   &    & IoT & 0.5	 & 1 & 4 &    & 1 & 1 &    \\
   
   & \multirow{-3}{*}{  GB} & eMBB & 4 & $R_{CRRM}$ & 3 & \multirow{-3}{*}{  10} & 0.014 & 8.82 & \multirow{-3}{*}{  315.1509} \\
  \cline{2-10}   
   &    & Voice & 0.032	 & 0.064 & 5 &    & 1 & 0.064 &    \\
   
   &    & IoT & 0.5	& 1 & 4 &    & 1 & 1 &    \\
   
   & \multirow{-3}{*}{  BG} & eMBB & 4 & $R_{CRRM}$ & 3 & \multirow{-3}{*}{  5} & 0.0106 & 6.678 & \multirow{-3}{*}{  252} \\
  \cline{2-10}
   &    & Voice & 0	 & 0.064 & 5 &    & 1 & 0 .064 &    \\
   
   &    & IoT & 0  & 1 & 4 &    & 0.7036 & 0.7037 &    \\
   
\multirow{-9}{*}{  630} & \multirow{-3}{*}{  BE} & eMBB & 0		 & $R_{CRRM}$ & 3 & \multirow{-3}{*}{  1} & 0.0014 & 0.882 & \multirow{-3}{*}{62.8491}\\
\hline
\end{tabular}
\end{table}

\begin{figure}[H]
	\centering
	\includegraphics[scale=0.8]{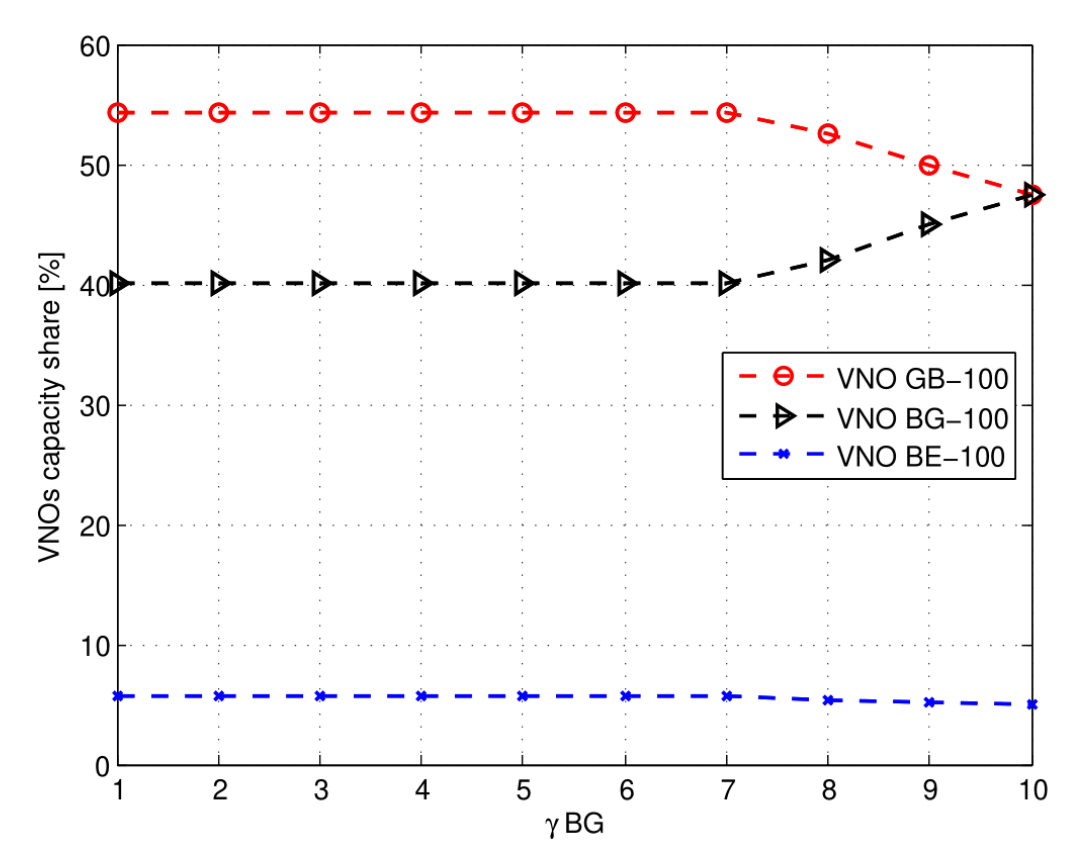}
	\caption{The influence that the weights of the VNOs have on the capacity sharing.}
	\label{fig15}
\end{figure}

$R^{vno}$ provided is basically the data rate share of each VNO, and $R^{srv}$ provided presents the data rate allocated to each service. The results shown in Table \ref{tab4} clearly states that the  $w^{usr}$  is highly dependent on the weights of both of the VNO and service however, the data rate depends on the $R^{srvMax}$.
By analyzing the results obtained previously, the following is observed:
\begin{itemize}
\item The weight values W for the users requesting both voice and IoT services are equal to the maximum value which is 1, and this depends on their priorities and the low speeds required,
\item The data rate provided for each service is guaranteed for all users while respecting the constraint linked to the data rate of the service set up by the model,
\item The total capacity allocated to each VNO is sufficient to serve all active users and also respects the second constraint, which perfectly reflects the reliability and performance of the model in terms of capacity sharing,
\item It is also observed that both of the voice and IoT services don’t use high data rate, as they just need a consistent data stream to function properly. As for the eMBB service it needs a marginally higher data rate, however, it differs for each VNO. ($\gamma_{BG}$)
\end{itemize}
It is taken for granted that the weight of VNO BG ($\gamma_{BG}$)  fluctuates between 1 and 10 and both of the GB and BE VNOs are fixed to their initial values to test the impact of weight tuning on the overall percentage of capacity held by the VNOs. Results for 100 users per VNO are illustrated in Figure \ref{fig15}. Pertaining to smaller values of  $\gamma_{BG}$, although the first assumption is to have equivalent levels of capacity sharing with VNO BE (given that the BE and BG VNOs have weights that are identical each other), it can be seen that VRRM maintains the values of BG at 40\%. This is VNO BG's guaranteed minimum capacity, which is substantially greater than VNO BE's shared capacity. By incrementing $\gamma_{BG}$  the capacity share stays at 40\% up $\gamma_{BG}$ equals seven (7), where the capacity shares of VNO GB and VNO BG continue to approach one another and eventually coincide at $\gamma_{BG}$ equals ten (10).

These findings show that all VNOs are satisfied with their SLAs separately within the physical infrastructure of InP, whereas all available capacity is shared 100\% between the three VNOs.

\subsection{The influence of the quality of the VNO radio link}

In this experiment we ran our model through different simulations where we fixed the total number of users to 300 user, and each VNO has the same amount of users (100 user per VNO) and we vary the max capacity of the radio link the network ($R_{CRRM}$) from -50\% to +50\% with adding 25\% each time of its original value of $R_{CRRM}$ which is set to 630 Mbps. The aim is to study its influence on the overall data rate allocated to the individual users of the three VNOs. The results of the variation of $R_{CRRM}$ on the weight of users $w^{usr}$ are presented in Tables \ref{tab5}, \ref{tab6} and \ref{tab7} for VNO1, VNO2, and VNO3 respectively. 

\begin{table}[H]
\centering
\caption{The influence of quality on the VNO 1 radio link.}
\label{tab5}
\begin{tabular}{llllll}
\hline
Service & -50\% $R_{CRRM}$ & -25\% $R_{CRRM}$ & $R_{CRRM}$ & +25\% $R_{CRRM}$ & +50\% $R_{CRRM}$ \\
\hline
Voice & 1 & 1 & 1 & 1 & 1 \\
Iot & 0.7857 & 1 & 1 & 1 & 1 \\
eMBB & 0.0127 & 0.0132 & 0.014 & 0.0144 & 0.0149 \\
\hline
\end{tabular}
\end{table}

\begin{table}[H]
\centering
\caption{. The influence of quality on the VNO 2 radio link.}
\label{tab6}
\begin{tabular}{llllll}
\hline
Service & -50\% $R_{CRRM}$ & -25\% $R_{CRRM}$ & $R_{CRRM}$ & +25\% $R_{CRRM}$ & +50\% $R_{CRRM}$ \\
\hline
Voice & 1 & 1 & 1 & 1 & 1 \\
Iot & 0.5 & 1 & 1 & 1 & 1 \\
eMBB & 0.0127 & 0.0097 & 0.0106 & 0.0112 & 0.0115 \\
\hline
\end{tabular}
\end{table}

\begin{table}[H]
\centering
\caption{The influence of quality on the VNO 3 radio link}
\label{tab7}
\begin{tabular}{llllll}
\hline
Service & -50\% $R_{CRRM}$ & -25\% $R_{CRRM}$ & $R_{CRRM}$ & +25\% $R_{CRRM}$ & +50\% $R_{CRRM}$ \\
\hline
Voice & 1 & 1 & 1 & 1 & 1 \\
Iot & 0.0786 & 0.4993 & 0.7037 & 0.908 & 1 \\
eMBB & 0.0003 & 0.0013 & 0.0014 & 0.0014 & 0.0015 \\
\hline
\end{tabular}
\end{table}

\begin{figure}[H]
	\centering
	\includegraphics[scale=0.8]{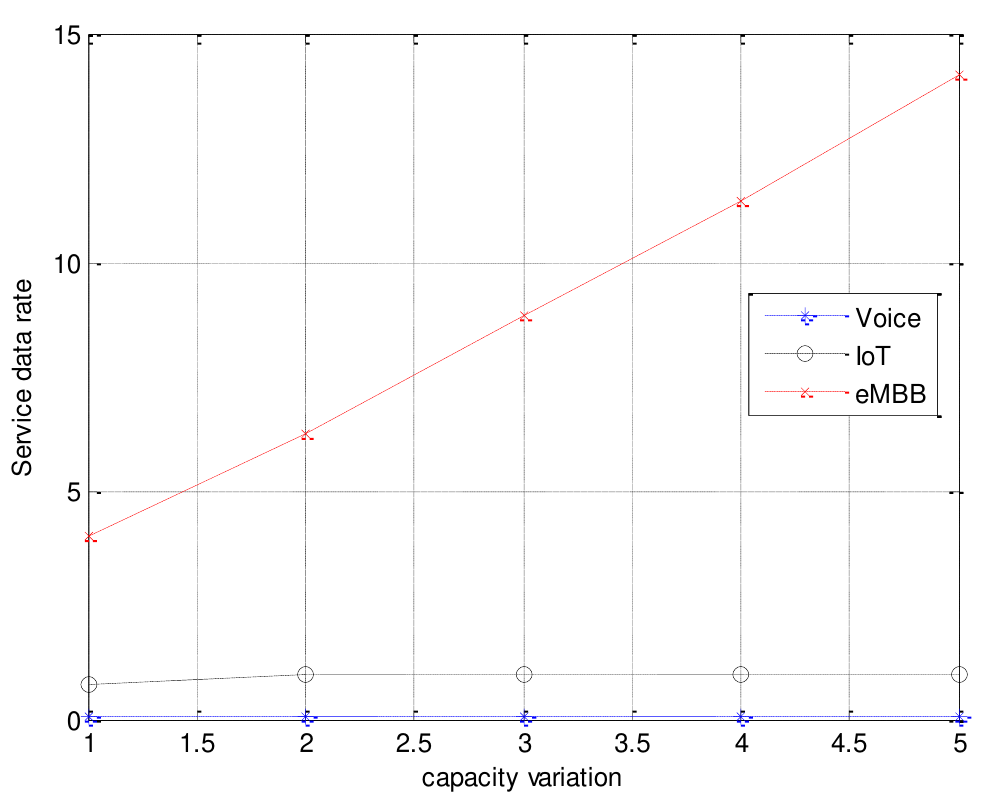}
	\caption{The influence of the quality of the VNO1 radio link.}
	\label{fig16}
\end{figure}

\begin{figure}[H]
	\centering
	\includegraphics[scale=0.8]{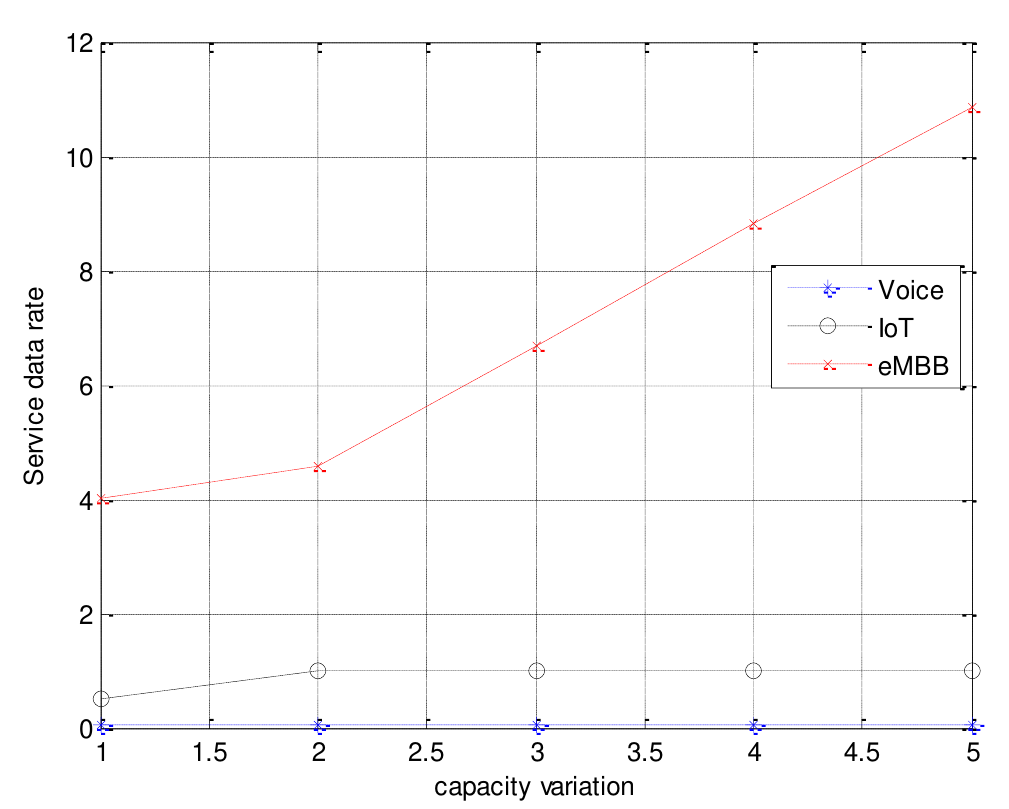}
	\caption{The influence of the quality of the VNO2 radio link.}
	\label{fig17}
\end{figure}

\begin{figure}[H]
	\centering
	\includegraphics[scale=0.8]{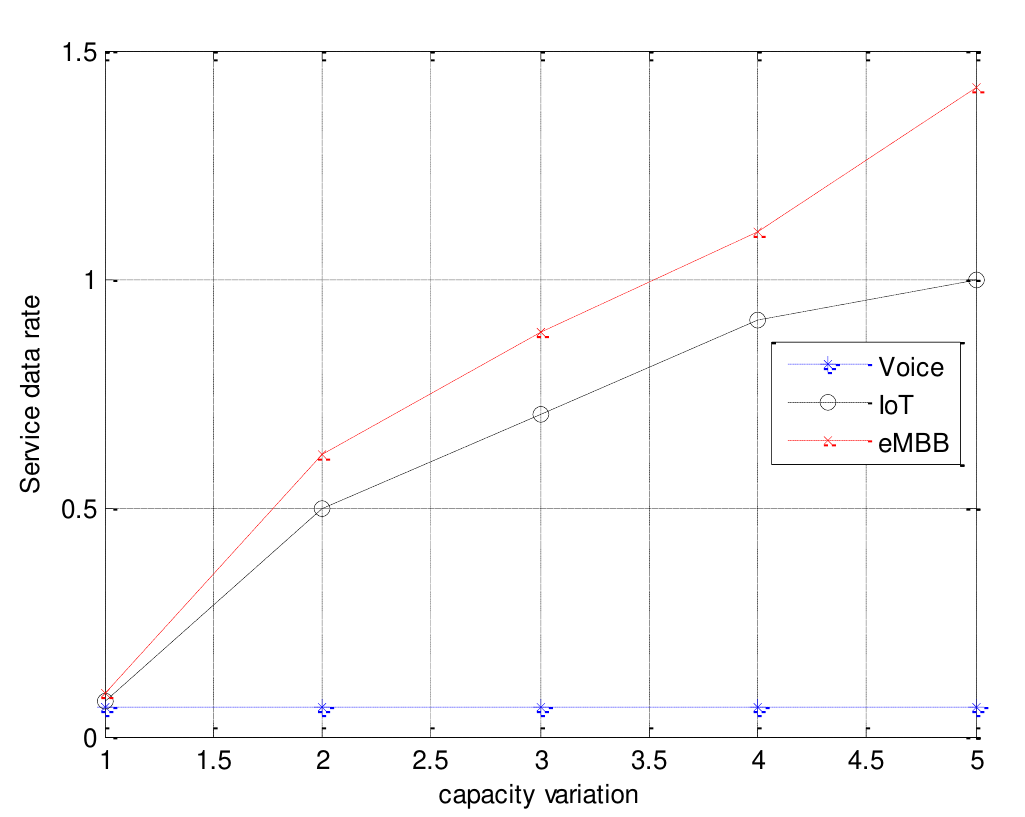}
	\caption{The influence of the quality of the VNO3 radio link.}
	\label{fig18}
\end{figure}

Figures \ref{fig16}, \ref{fig17}, and \ref{fig18} represents the obtained results of the variation of the maximum capacity shared $R_{CRRM}$ on the weight of service $R^{srv}$ for VNO1, VNO2 and VNO3 respectively.

We can observe from the results obtained above the following:
\begin{itemize}
    \item The data rate allocated to each of the voice service for the all three VNOs stayed somewhat consistent throughout all the different values of RCRRM,
	\item As for the IoT services, same can be said when it comes to VNO 1 and 2 but in VNO 3 it is noticed that by increasing the $R_{CRRM}$, the data rate increases to reach its maximum value at +50\% $R_{CRRM}$,
	\item Lastly, It is noticed that significant variance in the data rate of the eMBB services depending on the value of $R_{CRRM}$ especially in the third VNO. Also, the data rate is highly dependent on  $\sigma_v$ as we can see a decrease each time, we pass form a higher priority VNO to a lower one,
	\item The increase or decrease in the data rate assigned to each service follows a non-linear function which is justified by the use of the logarithm function present in the optimization model,
	\item For an increase or decrease of the overall capacity by 25\% or 50\%, the data rate of each service will be allocated according to the weight of the corresponding VNO, for example:
	\begin{itemize}
	    \item Increase by 25\%: the eMBB service of VNO1 increases by 28\%, VNO2 increases by 32\% and increases approximately by 25\% for VNO3,
	    \item Decrease of 50\%: the eMBB service of VNO1 decreases by 54\%, VNO2 decreases by 40\% and by 89\% for VNO3,
	    \item It has come to our attention that all of the VNOs are fully satisfied following the terms of their SLA contracts whatever the radio conditions,
	    \item The total available capacity is shared between the three VNOs.
	\end{itemize}
\end{itemize}

\subsection{The influence of the number users} 

We simulated different scenarios on our model similarly to the previous case however, this
time the $R_{CRRM}$ was fixed and we vary the number of users by adding 20\% of the max capacity of network. As the maximum number of users can’t go over 645 in totality, because the network is limited with the fixed value of the radio link (the $R_{CRRM}$ needs to be increased to increase the capacity of supported users), where each VNO has the same number of users each time. We have studied the effects on the allocated date rate to each individual user in the different services offered in the three VNOs. The results of the variation of the user number on the weight of user $w^{usr}$ are described in Tables \ref{tab8}, \ref{tab9}, and \ref{tab10} for VNO1, VNO2, and VNO3 respectively.

\begin{table}[H]
\centering
\caption{The influence of the number of users on VNO 1.}
\label{tab8}
\begin{tabular}{llllll}
\hline
Service & 20\% & 40\% & 60\% & 80\% & 100\% \\
\hline
Voice & 1 & 1 & 1 & 1 & 1 \\
Iot & 1 & 1 & 1 & 1 & 0.5 \\
eMBB & 0.04 & 0.016 & 0.001 & 0.0063 & 0.0063 \\
\hline
\end{tabular}
\end{table}

\begin{table}[H]
\centering
\caption{The influence of the number of users on VNO 2.}
\label{tab9}
\begin{tabular}{llllll}
\hline
Service & 20\% & 40\% & 60\% & 80\% & 100\% \\
\hline
Voice & 1 & 1 & 1 & 1 & 1 \\
Iot & 1 & 1 & 1 & 1 & 0.5 \\
eMBB & 0.03 & 0.0126 & 0.0077 & 0.0063 & 0.0063 \\
\hline
\end{tabular}
\end{table}

\begin{table}[H]
\centering
\caption{The influence of the number of users on VNO 3.}
\label{tab10}
\begin{tabular}{llllll}
\hline
Service & 20\% & 40\% & 60\% & 80\% & 100\% \\
\hline
Voice & 1 & 1 & 1 & 1 & 0.832 \\
Iot & 1 & 0.8367 & 0.5150 & 0.2356 & 0.0171 \\
eMBB & 0.004 & 0.0016 & 0.001 & 0.0004 & 0.00003 \\
\hline
\end{tabular}
\end{table}

\begin{figure}[H]
	\centering
	\includegraphics[scale=0.7]{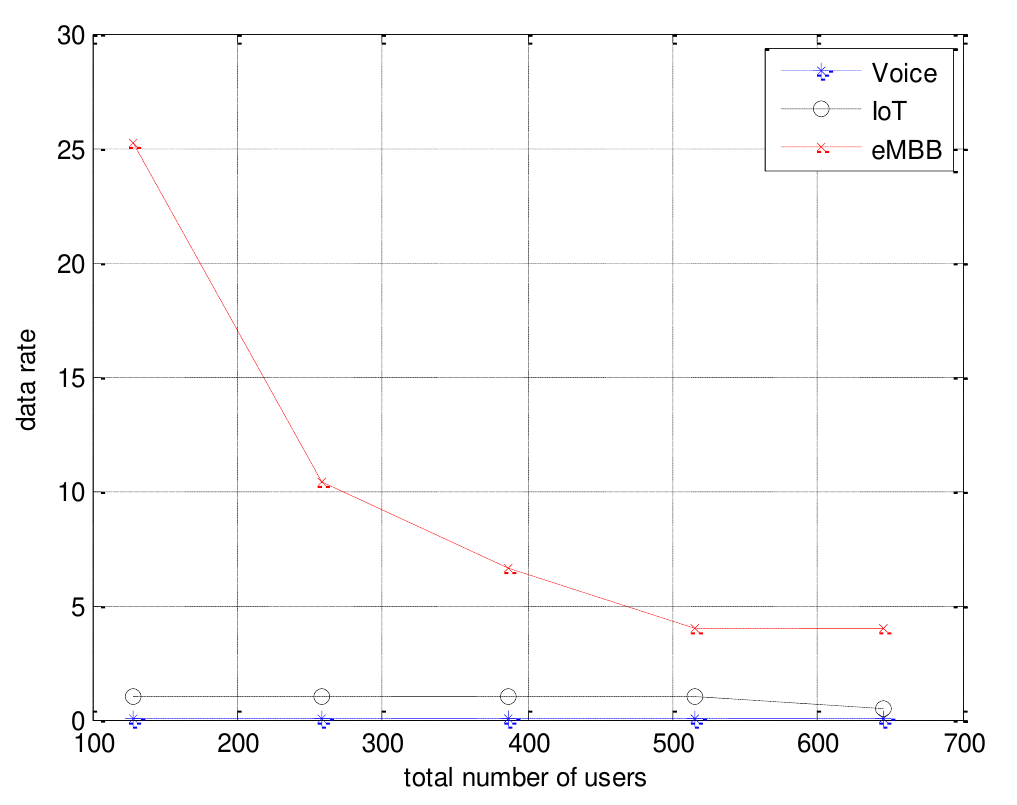}
	\caption{The influence of the number of users on the allocated data rate in VNO1.}
	\label{fig19}
\end{figure}

\begin{figure}[H]
	\centering
	\includegraphics[scale=0.7]{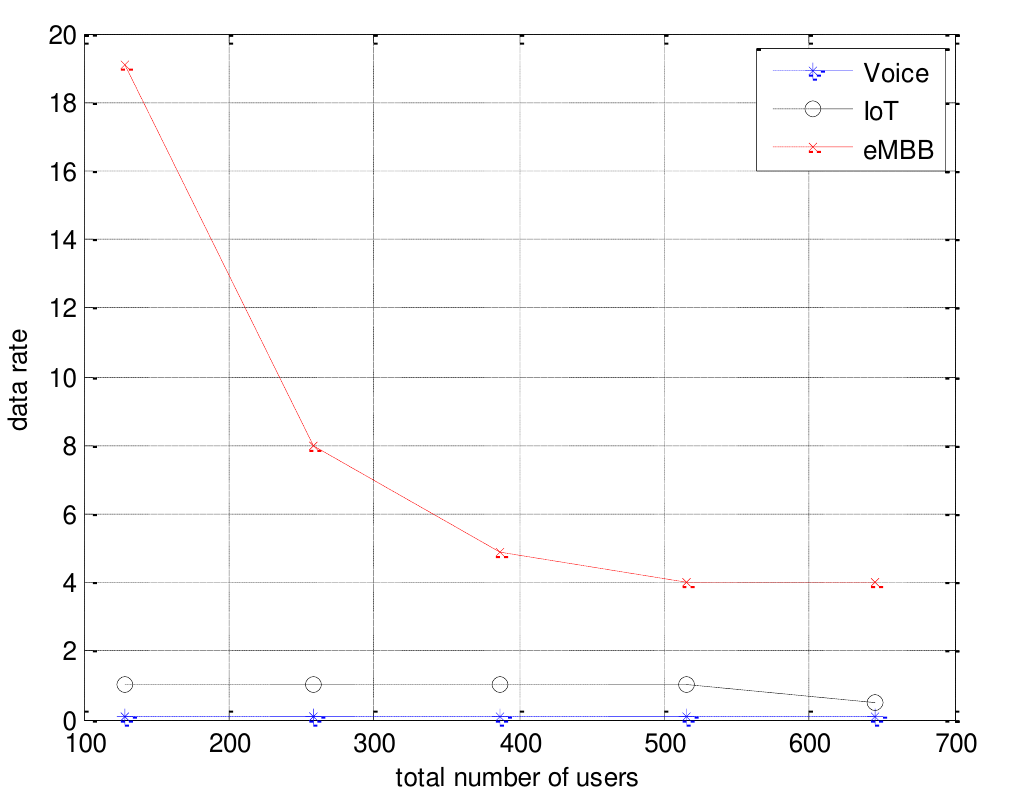}
	\caption{The influence of the number of users on the allocated data rate in VNO2.}
	\label{fig20}
\end{figure}

\begin{figure}[H]
	\centering
	\includegraphics[scale=0.7]{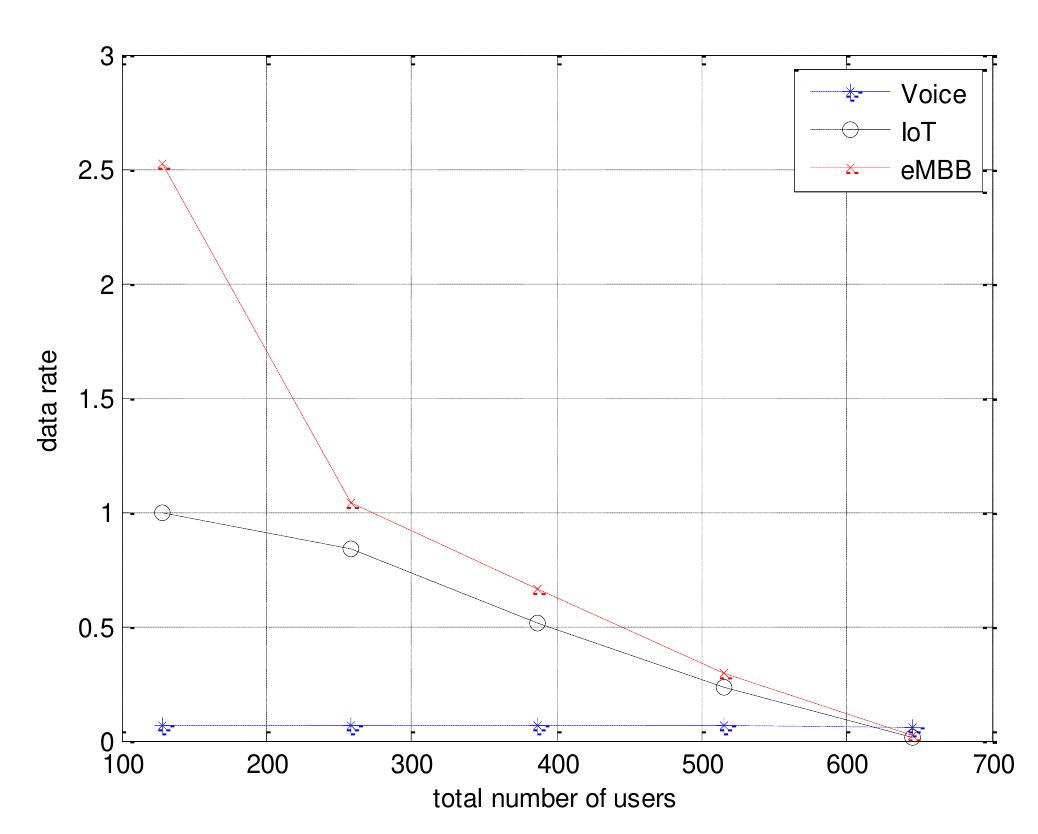}
	\caption{The influence of the number of users on the allocated data rate in VNO3.}
	\label{fig21}
\end{figure}

Figure \ref{fig19}, Figure \ref{fig20}, and Figure \ref{fig21} represents the obtained results of the variation of the user number on the weight of service $R^{srv}$ for VNO1, VNO2 and VNO3 respectively. 

It is noticed that a few changes occurred compared to the previous experiment:
\begin{itemize}
    \item The maximum number of users can’t go over $R_{CRRM}$ value in totality as the network is limited with the fixed value of the radio link (the $R_{CRRM}$ needs to be increased to increase the capacity of supported users),
    \item When it comes to the voice service, the data rate allocated to each of the three VNOs stayed consisting throughout. However, we noticed a minimal dip in the data rate in the case were the VNO3 was 100\% occupied.
    \item For the first two VNOs, the data rate allocated to the IoT service stayed at its maximum value of 1 Mbps, except when the VNO 3 were at 100\% of its user capacity it decreased to 0.5 Mbps,
    \item The data rate allocated to the eMBB in the three VNOs experienced a significant variance in values. The data rate decreases each the VNO has a lower weight, and the capacity of supported users get it maximum limit.
\end{itemize}

\section{Conclusion}

This paper discussed the  main features of the 5G which are SDN and NFV that enabled the ability of hardware virtualization and brought many advanced key functions to 5G network such as NS orchestration for managing slices in the network and infrastructure sharing.

In the second part, a set of optimization has been done on the proposed network model where we experimented, with varying the radio link throughput quality and the number of the user. We fixed the throughput and vary the number of users and verse-versa, to study the influence of each one on the allocated data rate. We conclude that by increasing the value of $R_{CRRM}$ it maximizes the data rate allocated to the users in each VNOs. To satisfy all the users of the network we need to set the $R_{CRRM}$. to an optimal value dictated by the use case.
It is noted that using the HS, all the VNOs are independently satisfied according to their SLA contracts and whatever the radio conditions. An example of 5G slicing architecture Matlab source code that satisfies our conditions can be found in the Github platform \footnote{\href{ https://github.com/usthbstar/5G\_HorizontalSlicing}{https://github.com/usthbstar/5G\_HorizontalSlicing}}

\bibliographystyle{unsrtnat}
\bibliography{5gSlice}  






\end{document}